\documentclass[11pt]{article}
\usepackage{verbatim,amsmath,amssymb}
\usepackage{epsfig,float,color}
\usepackage{epstopdf}
\usepackage{geometry}
\usepackage{setspace}
\usepackage{wrapfig}
\usepackage{hyperref}
\usepackage[utf8]{inputenc}
\usepackage{graphicx}
\usepackage{flushend}
\usepackage{tikz,pgfplots}
\usepackage[font={footnotesize}]{caption}
\usepackage[font={footnotesize}]{subcaption}
\usepackage{footnote}
\usepackage{multicol}
\usepackage{multirow}
\usepackage{enumitem}   
\usepackage{algorithm}%
\usepackage{algorithmicx}%
\usepackage{algpseudocode}%
\usepackage{soul}
\usepackage{framed}
\usepackage[titletoc,title]{appendix}
\usepackage{bm}
\usepackage{siunitx}
\usepackage[makeroom]{cancel} 
\usepackage{natbib}
\usepackage{hyperref}
\usepackage{bigints} 

\newenvironment{rcases}
{\left.\begin{aligned}}
	{\end{aligned}\right\rbrace}

\definecolor{darkblue}{rgb}{0,0,1}
\hypersetup{pdftex=true, colorlinks=true, breaklinks=true, linkcolor=blue, menucolor=blue, citecolor=black, urlcolor=blue}

\newcommand{\trr}[1]{{#1}^{\!\top}}

\newcommand{\inv}[1]{{#1}^{\text{-}1}}

\newcommand{\xpmt}[1]{\mathtt{#1}}
\newcommand{\xptt}[1]{\texttt{#1}}

\usepackage{tikz}
\usetikzlibrary{positioning, arrows.meta}
\tikzset{%
	myarrow/.style = {-Stealth, shorten >=5pt}
}

\usepackage{listings}
\definecolor{LightCyan}{rgb}{0.88,1,1}
\definecolor{mygreen}{RGB}{28,172,0} 
\definecolor{mylilas}{RGB}{170,55,241}
\newcommand{\PKcom}[1]{\textcolor{black}{#1}}

\begin{document}
	
	\begin{center}
		\Large{\bf{{\texttt{SoRoTop}: a hitchhiker's guide to topology optimization MATLAB code for design-dependent pneumatic-driven soft robots}}}\\
		
	\end{center}
	
	\begin{center}

		\large{Prabhat  Kumar\footnote{pkumar@mae.iith.ac.in}}
		\vspace{4mm}
		
		\small{\textit{Department of Mechanical and Aerospace Engineering, Indian Institute of Technology Hyderabad, 502285, India}}
			\vspace{4mm}
			
 Published\footnote{This pdf is the personal version of an article whose final publication is available at \href{https://link.springer.com/article/10.1007/s11081-023-09865-1}{Optimization and Engineering}}\,\,\,in \textit{Optimization and Engineering}, 
			\href{https://link.springer.com/article/10.1007/s11081-023-09865-1}{DOI:10.1007/s11081-023-09865-1} \\
			Submitted on 14~June 2023, Revised on 14~October 2023, Accepted on 14~October 2023

	\end{center}
	
	\vspace{1mm}
	\rule{\linewidth}{.15mm}
	{\bf Abstract:}
	  Demands for pneumatic-driven soft robots are constantly rising for various applications. However, they are often designed manually due  to the lack of systematic methods. Moreover, design-dependent characteristics of pneumatic actuation pose distinctive challenges. This paper provides a compact MATLAB code, named \texttt{SoRoTop}, and its various extensions for designing pneumatic-driven soft robots using topology optimization. The code uses the method of moving asymptotes as the optimizer and builds upon the approach initially presented in Kumar et al.(Struct Multidiscip Optim 61 (4): 1637–1655, 2020). The pneumatic load is modeled using Darcy's law with a conceptualized drainage term. Consistent nodal loads are determined from the resultant pressure field using the conventional finite element approach. The robust formulation is employed, i.e., the eroded and blueprint design descriptions are used. A min-max optimization problem is formulated using the output displacements of the eroded and blueprint designs. A volume constraint is imposed on the blueprint design, while the eroded design is used to apply a conceptualized strain energy constraint. The latter constraint aids in attaining optimized designs that can endure the applied load without compromising their performance. Sensitivities required for optimization are computed using the adjoint-variable method. The code is explained in detail, and various extensions are also presented. It is structured into pre-optimization, MMA optimization, and post-optimization operations, each of which is comprehensively detailed. The paper also illustrates the impact of load sensitivities on the optimized designs. \texttt{SoRoTop} is provided in Appendix~\ref{sec:sorotop} and is available with extensions in the supplementary material and publicly at \url{https://github.com/PrabhatIn/SoRoTop}. \\
	

	\vspace{-4mm}
	\rule{\linewidth}{.15mm}

\section{Introduction}
\begin{figure*}[h!]
	\centering
	\begin{subfigure}[t]{0.450\textwidth}
		\centering
		\includegraphics[scale=1]{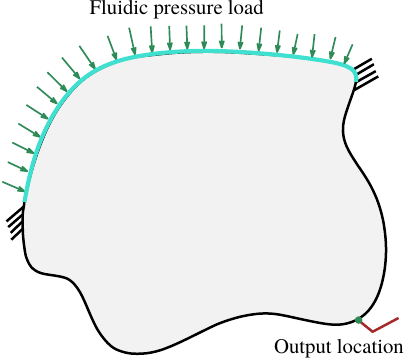}
		\caption{}
		\label{fig:Schematic1}
	\end{subfigure}
	\quad \quad
	\begin{subfigure}[t]{0.45\textwidth}
		\centering
		\includegraphics[scale=1]{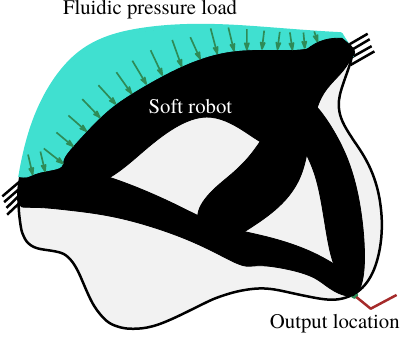}
		\caption{}
		\label{fig:Schematic2}
	\end{subfigure}
	\caption{A schematic diagram for a pressure-driven soft robot. (\subref{fig:Schematic1}) Design domain. A set of arrows indicates the fluidic pressure load. Fixed boundary conditions are also depicted.  (\subref{fig:Schematic2}) A representative solution. One notices that pressure } \label{fig:Schematic}
\end{figure*}
Robotics has gained unprecedented attention and development across various applications, including industry, academia, household chores, official tasks, and defense applications, to name a few. Additionally, there is a growing interest in ``soft robotics'' due to its unique advantages over traditional rigid robots that rely on solid linkage mechanisms. Soft robots are designed using materials with Young's modulus ranging from kilopascals to megapascals.
These robots perform tasks by utilizing the deformation of their flexible bodies, making them well-suited for various purposes, such as human interaction in unstructured and dynamic environments,  fragile objects handling, fruit and vegetable picking and placing, and achieving high power-to-weight ratios~\citep{kumar2022towards,xavier2022soft}. Thus, nowadays, they find various applications in, e.g., gripping~\citep{xie2020octopus,pinskier2022automated}, sensing~\citep{zhao2016optoelectronically}, invasive surgery~\citep{hu2018steerable}, rehabilitation~\citep{polygerinos2013towards}, handling soft and fragile objects~\citep{shintake2018soft,deimel2013compliant}, and human-robot interaction~\citep{deimel2013compliant}, etc. These robots can be classified based on the ways they require actuation. For example,  fluidic pressure-driven actuation~\citep{kumar2020topology,kumar2022towards,gorissen2017elastic,kumar2022topological,de2020topology,lu2021topology,vasista2012design}, cable-driven actuation~\citep{chen2018topology}, electroactive polymer-based actuation~\citep{pourazadi2019investigation} and shape-memory material-based actuation~\citep{jin2016soft}, etc. Among these, fluidic pressure (pneumatic)-driven robots are more sought after because of their lightweight, quick response, and low-cost components~\citep{kumar2022towards,xavier2022soft}. However,  due to a lack of systematic approaches, they are designed manually using heuristic methods that heavily depend on designers' knowledge, experience, and expertise. Such heuristic methods often require numerous resources/iterations to achieve the desired designs; thus, they introduce limitations and may not be efficient.

Characteristic features of soft robots resemble those of compliant mechanisms that perform their tasks utilizing the motion obtained from their flexible members~\citep{kumar2022towards}. Thus, topology optimization methods for designing pneumatically actuated compliant mechanisms (CMs) can be one of the promising ways to design soft robots~\citep{kumar2022topological,kumar2022towards}. Topology optimization (TO), a computational technique, determines where to place material and where to make holes (void) within a given design domain to achieve optimum performance. The design domain is typically parameterized using finite elements (FEs), and each element is assigned a design variable~$\rho\in[0,\,1]$. $\rho_i=1$ indicates material or solid phase while $\rho_i=0$ denotes void state of element~$i$. The optimized designs are expected to contain FEs with $\rho=1$. However, the TO problem is often relaxed in practical applications, leading to optimized designs that may include elements with $0<\rho<1$~\citep{sigmund2013topology}. To mitigate this and guide solutions toward 0-1, the Heaviside projection-based filter can be employed~\citep{wang2011projection}.

A design-dependent load changes its magnitude, direction, and/or location as topology advances~\citep{kumar2020topology}. Therefore, such loads pose several unique challenges within a TO setting~\citep{kumar2020topology,hammer2000topology,kumar2023TOPress}. These challenges encompass:
(a) locating the boundary to apply the pneumatic load, which involves relating the pressure field to the employed design variables, (b) converting the fluidic pressure field into consistent nodal loads, and (c) effectively evaluating the load sensitivities. These challenges can become more prominent when designing pneumatic-activated soft robots/compliant mechanisms~\citep{kumar2020topology,kumar2021topology}. Fig.~\ref{fig:Schematic1} displays a schematic diagram, and Fig.~\ref{fig:Schematic2} depicts its representative optimized soft robots design. One notes that the pressure boundary changes, and the load's direction, magnitude, and location get updated. Typically, soft robots experience large deformation while performing their tasks. Elements/regions with low stiffness deform significantly; thus, they pose an additional set of challenges~\citep{wang2014interpolation,kumar2021topologyDTU}. In addition, for capturing the design-dependent nature of the pneumatic loads with large deformation, one needs to include the follower force concepts within the formulation~\citep{kumar2022topological}. Moreover, when designing mechanisms, there may be instances where members come into contact with each other, further complicating the formulation~\citep{kumar2019computational}. Given the various nonlinearities and complexities associated with these challenges, it is worth noting that the discussion in this paper is confined to a linear and standard finite element setting, and detailed discussions of each nonlinearity are beyond the scope of the provided MATLAB code.

\citet{hammer2000topology} were the first to present fluid pressure load within the TO setting for designing loadbearing structures. Since then, many approaches have been developed to explicitly design loadbearing structures by minimizing their compliance; see Refs.~\citep{kumar2020topology,picelli2019topology,kumar2021topology} and related reference therein for a detailed overview. \citet{hiller2011automatic} employed the evolutionary TO method to design soft robots, while \citet{chen2018topology} utilized level-set TO to create a soft cable-driven gripper. \citet{zhang2018design} designed a soft pneumatic gripper. However, none of the methods  in these studies~\citep{hiller2011automatic,chen2018topology,zhang2018design} considered for the design-dependent characteristics of the pneumatic loads in their approaches. Recently, \citet{pinskier2022automated}  presented TO methods for 3D multi-material soft grippers building upon the first introduced in~\citep{kumar2020topology,kumar2021topology}. In addition, to date, only a limited number of TO methods exist to design fluidic pressure-actuated compliant mechanisms considering the design-dependent nature of the loads in 2D~\citep{kumar2020topology,kumar2022topological,panganiban2010topology,vasista2012design, de2020topology,lu2021topology}  and 3D~\citep{panganiban2010topology,kumar2021topology} settings. Given the various challenges and complexities involved in these approaches, newcomers, students and even experienced researchers in this field may face difficulties in developing new approaches or using existing ones. Therefore, to facilitate their learning and research, we need a freely available systematic MATLAB code for designing pneumatically actuated soft robots.

Providing open codes and sub-routines within research communities is highly appreciated, as they greatly facilitate the hands-on experience and efficient learning of the (new) concepts. The codes also facilitate reproducibility and provide confidence in the outcomes.
In the current state-of-the-art of TO, readers can find numerous MATLAB codes for different applications such as~\citep{kumar2022honeytop90,sigmund200199,suresh2010199,andreassen2011efficient,ferrari2021topology,picelli2021101,homayouni20212d,gao2019concurrent,ali2022toward,alexandersen2023detailed,kumar2023TOPress}.  For an in-depth review of such educational codes, refer to~\cite{wang2021comprehensive}. Despite the increasing interest in automated methods for designing pneumatic-driven soft robots, no open-source codes are currently available. Therefore, there is a significant need for open-source MATLAB codes in this field to support the growing research interest and educational needs.

This paper aims to provide a MATLAB code for systematically designing pneumatic-driven soft robots to ease the learning  in this field. The code is also expected to furnish a suitable platform for researchers to develop, explore, and extend for designing various soft robots for different applications.

The presented code,  named $\texttt{SoRoTop}$, is developed using the approach initially introduced in~\cite{kumar2020topology} and  $\xpmt{TOPress}$ MATLAB code~\citep{kumar2023TOPress}.  The density-based TO setting is considered. We use the robust formulation~\citep{wang2011projection} with blueprint (intermediate) and eroded designs, i.e., the optimized mechanisms
are robust regarding over-etching (see Fig.~\ref{fig:Fil_ero}). Over-etching involves exposing the substrate to an etching process longer than necessary to remove excess material. The method of moving asymptotes (MMA, written in 1999 and updated in the 2002 version)~\citep{svanberg1987} is used as the optimizer, allowing users to extend the code with additional physics and multiple constraints readily. A noteworthy aspect of the code is its consideration of load sensitivities. Load sensitivities exist due to the design-dependent nature of the load as discussed in~\citep{kumar2020topology,kumar2023TOPress}. The code also demonstrates the importance and effects of load sensitivities on the optimized mechanisms.

The remainder of the paper is structured as follows. Sec.~\ref{sec:PresureLoadmodeling} models pressure load as a design-dependent force and evaluates the consistent nodal forces. Sec.~\ref{sec:TopologyOptForm} describes topology optimization formulation, wherein the robust optimization scheme is described and sensitivity analysis is performed. MATLAB implementation for $\xpmt{SoRoTop}$ code is described in detail in Sec.~\ref{sec:MATLABImple}. Numerical results using $\xpmt{SoRoTop}$ are presented in Sec.~\ref{sec:ResultsDiscussion}. Various extensions of $\xpmt{SoRoTop}$ are provided for different pneumatically actuated soft mechanisms. The section also provides discussions on these results. Lastly, concluding remarks are presented in Sec.~\ref{sec:conclusions}. 

\section{Pressure load modeling and nodal forces evaluation}\label{sec:PresureLoadmodeling}
We confine ourselves to density-based TO~\citep{sigmund2013topology} herein. Thus, each element is assigned a design (density) variable considered constant within the element. Typically, optimization starts by initializing the element's density equal to the allowed element volume fraction, $\frac{V^*}{{nel}}$. $V^*$ represents the total volume fraction, and ${nel}$ indicates the total number of square bi-linear finite elements to parameterize the design domain. Next, we briefly describe the modeling of pressure load and evaluation of consistent nodal force evaluation for the sake of completeness. One may refer to~\cite{kumar2020topology} for a detailed description.
\subsection{Design-dependent pressure load modeling}\label{Sec:Design-dependent_PL}
As optimization progresses, material phases of elements change, and elements can be considered a porous medium. In addition, for the given design problem, the fluidic pressure boundary conditions are known \textit{a priori}. Therefore, per~\cite{kumar2020topology,kumar2021topology},  Darcy law is a natural and appropriate choice to model pressure load or to establish the relationship between pressure load and the design variables~. The Darcy flux $\bm{q}$ is determined as
\begin{equation}\label{Eq:Darcyflux}
	\bm{q} = -\frac{\kappa}{\mu}\nabla p = -K(\bar{\bm{\rho}}) \nabla p,
\end{equation}
where $\nabla p$, $\kappa$, and $\mu$ represent the pressure gradient, permeability of the medium, and fluid viscosity, respectively. $\bar{\bm{\rho}}$ is the physical variable, and  $K(\bar{\bm{\rho}})$ is termed flow coefficient. The latter is related to the former for element~$e$ as
\begin{equation}\label{Eq:Flowcoefficient}
	K(\bar{\rho_e}) = K_v\left(1-(1-\epsilon) \mathcal{H}(\bar{{\rho_e}},\,\beta_\kappa,\,\eta_\kappa)\right),
\end{equation} 
where $\epsilon = \frac{K_s}{K_v}$ is the flow contrast. $K_v$ and $K_s$ are the flow coefficients of the void and solid phases of an element.
\begin{equation} \mathcal{H}(\bar{{\rho_e}},\,\beta_\kappa,\,\eta_\kappa) = \frac{\tanh{\left(\beta_\kappa\eta_\kappa\right)}+\tanh{\left(\beta_\kappa(\bar{\rho}_e - \eta_\kappa)\right)}}{\tanh{\left(\beta_\kappa \eta_\kappa\right)}+\tanh{\left(\beta_\kappa(1 - \eta_\kappa)\right)}}
\end{equation}
is a smooth Heaviside function. $\eta_\kappa$ defines transition point, whereas $\beta_\kappa$ indicates steepness for the flow coefficient. We set $K_v =1$, and $\epsilon = \SI{1e-7}{}$~\citep{kumar2021topology,kumar2023TOPress}. Solving Eq.~\ref{Eq:Darcyflux} may not give the realistic pressure variation as demonstrated in  ~\cite{kumar2021topology,kumar2022improved,kumar2023TOPress}. Thus, the additional volumetric drainage term $Q_\text{drain}$  is conceptualized and included in the Darcy law. Now, the  balance equation can be written as
\begin{equation}\label{Eq:stateequation}
	\nabla\cdot\bm{q} - Q_\text{drain} = \nabla \cdot \left(K(\bar{\bm{\rho}}) \nabla p\right)+  Q_\text{drain}=0,
\end{equation}
where ${Q}_\text{drain} = -D(\bar{\rho_e}) (p - p_{\text{ext}})$ with
\begin{equation}\label{Eq:drainageterm}
	D(\bar{\rho_e}) =  D_{\text{s}}\mathcal{H}(\bar{{\rho_e}},\,\beta_d,\,\eta_d),
\end{equation}
where $\left\{\eta_\text{d},\,\beta_\text{d}\right\}$ are the drainage parameters. $\eta_\text{d}$ and $\beta_\text{d}$ define the transition and steepness for the drainage term, respectively. $p_{\text{ext}}$ is the external pressure field. Per~\cite{kumar2020topology}, $D_{\text{s}} = \left(\frac{\ln{r}}{\Delta s}\right)^2 K_\text{s}$, where $r = \frac{p|_{\Delta s}}{p_\text{in}}$. $p|_{\Delta s}$ is the pressure at $\Delta s$, a penetration parameter set to the width/height of a few FEs~\citep{kumar2020topology}. In general, $r\in[0.001 \,\,  0.1]$ is set. Per~\cite{kumar2023TOPress}, user-defined parameters are reduced by considering $\eta_d = \eta_k = \eta_f$ and $\beta_d = \beta_k =\beta_f$ for the provided code, \texttt{SoRotop}. We indicate $\eta_f$ and $\beta_f$ in $\texttt{SoRotop}$ by $\texttt{etaf}$ and $\texttt{betaf}$, respectively.

Equation~\ref{Eq:stateequation} is solved using the standard finite element method, that transpires to~\citep{kumar2020topology,kumar2023TOPress}
\begin{equation} \label{Eq:FEAnumint}
	\begin{aligned}
		\mathbf{K}_\text{p}^e \mathbf{p}_e+ \mathbf{K}^e_\text{Dp} \mathbf{p}_e = \mathbf{A}_e \mathbf{p}_e = \mathbf{0},
	\end{aligned}
\end{equation}
where $\mathbf{\mathbf{K}}_\text{p}^e$ is the element flow matrix due to Darcy law, whereas $\mathbf{K}^e_\text{Dp}$ appears due to the drainage term. $\mathbf{A}_e$ indicates the overall element flow matrix. With the global flow matrix $\mathbf{A}$ and global pressure field $\mathbf{p}$, Eq.~\ref{Eq:FEAnumint} is written as
\begin{equation}\label{Eq:PDEsolutionpressure}
	\mathbf{Ap} = \mathbf{0}.
\end{equation}
One solves Eq.~\ref{Eq:PDEsolutionpressure} using the given pressure boundary conditions to obtain the overall pressure field $\mathbf{p}$.  $\mathbf{A}$ and $\mathbf{p}$ are sub-blocked into \textit{free} and  \textit{prescribed}, denoted using subscripts \textit{f} and \textit{p}, respectively; thus, Eq.~\ref{Eq:PDEsolutionpressure} can be written as
\begin{equation}\label{Eq:partEq6}
	\begin{bmatrix}
		\mathbf{A}_{ff} & \mathbf{A}_{fp} \\
		\mathbf{A}_{fp}^\top & \mathbf{A}_{pp} 
	\end{bmatrix}  
	\begin{bmatrix}
		\mathbf{p}_f \\
		\mathbf{p}_p 
	\end{bmatrix} = \begin{bmatrix}
		\mathbf{0} \\
		\mathbf{0}
	\end{bmatrix}.
\end{equation}
The first row of Eq.~\ref{Eq:partEq6} yields $\mathbf{p}_f = \mathbf{A}_{ff}^{-1} \mathbf{A}_{fp}\mathbf{p}_p$; thus, the pressure field $\mathbf{p}$ is determined. 

\subsection{Nodal forces evaluation}
The nodal forces are obtained from the pressure field as~\citep{kumar2020topology}
\begin{equation}\label{Eq:pressuretoload}
	\bm{b} \text{d}V = -\nabla p \text{d}V,
\end{equation}
where $\text{d}V$ is the infinitesimal volume and $\bm{b}$ is the body force~\citep{kumar2020topology}. Now, using the standard FE method, one writes at the elemental level
\begin{equation}\label{Eq:Forcepressureconversion}
	\mathbf{F}_e = -\left[\int_{\mathrm{\Omega}_e} \trr{\mathbf{N}}_\mathbf{u} \mathbf{B}_p  d {V}\right]\, \mathbf{p}_e = -\mathbf{T}_e\,\mathbf{p}_e,
\end{equation}
where $\mathbf{N}_\mathbf{u} = [N_1\mathbf{I},\, N_2\mathbf{I},\,N_3\mathbf{I},\,N_4\mathbf{I}]$, with $N_1,\,N_2,\,N_3,\,N_4$ are the bi-linear shape functions, $\mathbf{B}^\top_\text{p} = \nabla\mathbf{N}_\text{p}$ with $\mathbf{N}_\text{p} = [N_1,\,N_2,\,N_3,\,N_4]^\top$ and $\mathbf{I}$ is the identity matrix in $\mathcal{R}^2$. $\mathbf{T}_e$ is called the transformation matrix. $\mathbf{p}_e$ is the elemental pressure vector, i.e., $\mathbf{p}_e$ = $[p_1,\,p_2,\,p_3,\,p_4]^\top$. In the global sense, Eq.~\ref{Eq:Forcepressureconversion} transpire to
\begin{equation}\label{Eq:nodalforce}
	\mathbf{F} = -\mathbf{T}\mathbf{p},
\end{equation}
where $\mathbf{F}$ and $\mathbf{T}$ are the global force vector and transformation matrix, respectively. For the given problem, one uses Eq.~\ref{Eq:PDEsolutionpressure} to determine the overall pressure field in view of the given pressure boundary conditions, whereas Eq.~\ref{Eq:nodalforce} is used to determine the global nodal forces from the obtained pressure field $\mathbf{p}$.
\begin{figure*}[h!]
	\centering
	\includegraphics[scale=1]{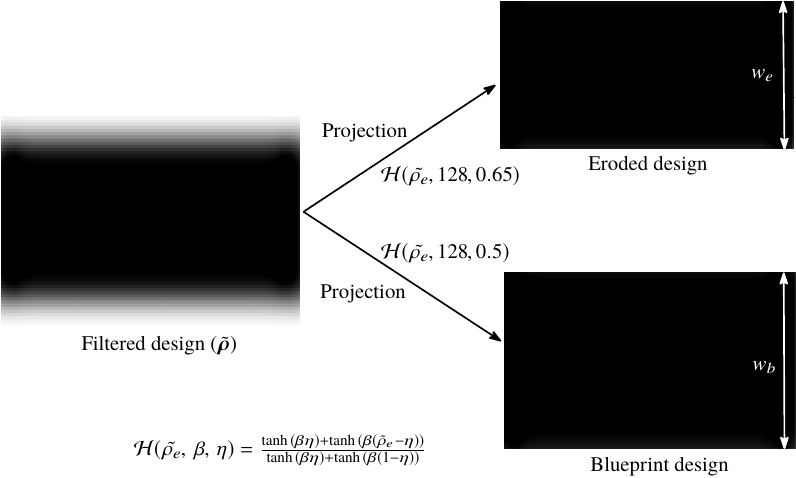}
	\caption{A schematic for filtered, eroded and blueprint designs. $w_e< w_b$, where $w_e$ and $w_b$ represent the widths  of the eroded  ($\eta = 0.65$) and  blueprint designs ($\eta = 0.5$), respectively.}
	\label{fig:Fil_ero}
\end{figure*}
\section{Topology optimization formulation}\label{sec:TopologyOptForm}
We use the modified SIMP formulation~\citep{sigmund2013topology} to interpolate the Young's modulus defined for element~$i$ as,
\begin{equation}\label{Eq:SIMP}
	\mathrm{E}_i = \mathrm{E}_\mathrm{min} + \bar{\rho}^p_i(\mathrm{E}_1-\mathrm{E}_\mathrm{min}),
\end{equation}
where $E_1$ and $E_\text{min}$ are Young's moduli of the solid and void phases of an element, respectively, and $E_i$ is interpolated Young's modulus of element~$i$. $E_\text{min}$ is assigned to the void elements to avoid the singularity of the structural stiffness matrix during finite element solve. $p$ is the SIMP parameter. $\bar{\rho}$, the physical design variable of element~$i$, is determined as
\begin{equation}\label{Eq:projectionFilt}
	\bar{\rho}_i = \frac{\tanh(\beta \eta) + \tanh(\beta(\tilde{\rho}_i-\eta))} {\tanh{\left(\beta \eta\right)}+\tanh{\left(\beta(1 - \eta)\right)}} = \mathcal{H}(\tilde{{\rho_i}},\,\beta,\,\eta),
\end{equation}
where $\eta$ indicates the transition point and  $\beta \in (0,\,\infty)$ represents the steepness parameter, for values greater than 0, it reduces the number of gray elements. To achieve binary optimized solutions, $\beta$ is increased towards infinity. However, for $\beta \to \infty$, $	\frac{\partial \bar{\rho}_i}{\partial \tilde{\rho}_i} \to 0$, i.e., derivatives of the objective become very small; thus jeopardize the optimization progress.  The choice of starting $\beta$ can vary; initializing with a higher number hinders the optimization search behavior due to the abovementioned reasons. Thus, we start with $\beta =1$ and increase it in a continuation fashion till $\beta_{\max} = 128$. The derivative of $\bar{\bm{\rho}}$ with respect to the $\tilde{\rho}$ can be determined as
\begin{equation}\label{Eq:der_PHYVariable_filvariable}
	\frac{\partial \bar{\rho}_i}{\partial \tilde{\rho}_i} = \beta\frac{1-\tanh(\beta(\tilde{\rho}-\eta))^2}{(\tanh{\left(\beta \eta\right)}+\tanh{\left(\beta(1 - \eta)\right)})^2},
\end{equation}
$\tilde{\rho}$ denotes the filtered design variable for element~$i$. Mathematically, we evaluate $\tilde{\rho}$ as
\begin{equation} \label{Eq:density_filter}
	\tilde{\rho}_i = \frac{\displaystyle\sum_{j=1}^{{nel}}\rho_j \mathrm{v}_j \mathrm{w}(\mathbf{x}_i,\mathbf{x}_j)}{\displaystyle\sum_{j=1}^{{nel}} \mathrm{v}_j \mathrm{w}(\mathbf{x}_i,\mathbf{x}_j) } \; ,
\end{equation}
where $\mathrm{w}(\mathbf{x}_i,\mathbf{x}_j)$ = $\mathrm{max}  \left(0 \; , \; 1-\frac{\| \mathrm{\mathbf{x}}_i - \mathrm{\mathbf{x}}_j \|}{\mathrm{r}_\mathrm{fill}} \right)$ \citep{bruns2001}. The filter radius is indicated by $\mathrm{r}_\mathrm{fill}$. $\mathrm{v}_j$ represents volume of element~$j$. $\mathrm{w}(\mathrm{x}_i,\mathrm{x}_j)$ is determined at the beginning of the optimization, as it does not alter with TO iterations. It is stored in a matrix  $\mathbf{Hs}$ in the following manner:
\begin{equation}\label{Eq:HSvalue}
	\mathrm{Hs}_{i,j} = \frac{\mathrm{v}_j \: \mathrm{w}(\mathrm{\mathbf{x}}_i,\mathrm{\mathbf{x}}_j)}{{\displaystyle\sum_{j=1}^{{nel}} \mathrm{v}_j \mathrm{w}(\mathrm{\mathbf{x}}_k,\mathrm{\mathbf{x}}_j) }}.
\end{equation}
Finally, the filtering process is performed as  $\bm{\tilde{\rho}}=\mathbf{Hs}\bm{\rho}$, and one determines its derivative with respect to $\bm{\rho}$ as
\begin{equation}\label{Eq:der_filVariable_ovariable}
	\frac{\partial \bm{\tilde{\rho}}}{\partial \bm{\rho}}=\mathbf{Hs}^\top.
\end{equation}
One may also use \texttt{infilter} MATLAB function for performing the filtering operations~\citep{ferrari2021topology,kumar2023TOPress}.

\subsection{Optimization problem formulation}

The optimized CMs obtained using topology optimization typically contain single-node connections~\citep{sigmund1997design,yin2003design}. These connections show artificial stiffness, permit load transfer with zero stiffness, and appear due to deficiencies in FE-analysis with quadrilateral elements. CMs with such hinges are difficult to realize as, in reality, compliant hinges will always have finite rotational stiffness. Numerous methods have been proposed to circumvent the introduction of point connections in the optimized CMs~\citep{yin2003design,poulsen2003new,saxena2007honeycomb,wang2011projection,singh2020topology}. Herein, we use the robust formulation~\citep{wang2011projection} with respect to the blueprint and eroded designs~\citep{kumar2022topological}. We replace $\eta$ by $0.5+\Delta\eta$ and $0.5$ in Eq.~\ref{Eq:projectionFilt} for determining the eroded and blueprint variables, respectively~(Fig.~\ref{fig:Fil_ero} cf.~\cite{wang2011projection}). $\Delta\eta \in [0,\,0.5]$ is a user-defined parameter. 

As mentioned before, the characteristics of soft robots (monolithic designs with no rigid joints) are similar to CMs~\citep{kumar2022towards}. TO of CMs for achieving maximum output displacements with the design-dependent behavior of the pneumatic loads can be one of the promising directions to design soft robots~\citep{kumar2022towards,de2020topology,pinskier2022automated}. A min-max, non-smooth optimization problem is formulated using the output deformation of the blueprint and eroded designs. The given volume constraint is applied to the blueprint design. This indicates that the optimized result is robust with respect to the over-etching~\citep{wang2011projection,fernandez2020imposing}. The optimized mechanisms are expected to sustain under the applied loads~\citep{saxena2000optimal}; therefore, a strain energy constraint is conceptualized and applied to the eroded design. The constraint aids in attaining optimized designs that can endure the applied load without compromising their performance. The optimization problem is solved using the method of moving asymptotes (MMA, cf.~\cite{svanberg1987}). Mathematically, the optimization problem is written as

\begin{equation}\label{Eq:Optimizationequation}
	\begin{rcases}
		\underset{\bar{\bm{\rho}}(\tilde{\bm{\rho}}(\bm{\rho}))}{\text{min}:}
		f_0 = \max_{r}\, {u^\text{out}_r} = \max_{r}\left\{\bm{l}^\top \mathbf{u}_r\right\}|_{r =e,\,b}\\
		\text{such that:} \\
		\,{^1}{\bm{\lambda}_r}:\,\, \mathbf{A}_r\mathbf{p}_r = \mathbf{0 }\\
		\,{^2}\bm{\lambda}_r:\,\,  \mathbf{K}_r\mathbf{u}_r = \mathbf{F}_r = -\mathbf{T} \mathbf{p}_r\\
		\,{\Lambda}_b:\,\, \text{g}_1 = \frac{V_b}{V^*}-1\le 0\\
		\,{\Lambda}_e:\,\,  \text{g}_2= \frac{SE_e}{SE^*}-1 \le 0\\
		\, 0\le \rho_i,\,\tilde{\rho}_i,\,\bar{\rho}_i\le 1 \,\,\forall i
	\end{rcases},
\end{equation}
where quantities with subscript $e$ and $b$ are related to the eroded and intermediate/blueprint designs, respectively~\citep{wang2011projection,kumar2022topological}. $\mathbf{u}_r$ is the global displacement vector, whereas $\mathbf{K}_r$ indicates the global stiffness matrix. $u^\text{out}_r$ denotes the output displacement of the mechanisms in the desired directions. $\bm{l}$ is a vector with all zeros except the entry corresponding to the output degree of freedom set to one. $V^*$ and $V_b$  are the blueprint design's permitted and current volume fractions, respectively. $SE_e$ and $SE^*$ are the eroded design's current and defined strain energy, respectively. We choose $SE^*$  at the initial optimization stage so that the optimized mechanism can sustain the applied load and deliver high performance. The  $SE^*$ and g$_2$ are determined as
\begin{algorithm}
	\caption{Calculation of $SE^*$ and $g_2$ formulation}\label{Alg:g2}
	\begin{algorithmic} 
		\Require $loop,\, \mathbf{u}_e,\, \mathbf{K}_e,\,SE_e = \frac{1}{2}\mathbf{u}_e^\top\mathbf{K}_e \mathbf{u}_e$
		\If{$loop==1$}             
		\State $SE^* = S_f\times SE_e$ 
		\EndIf
		\State $g_2 = \frac{SE_e}{SE^*}-1\le 0$
	\end{algorithmic}
\end{algorithm}

\noindent where $S_f$, a user-defined strain energy fraction, is selected to obtain a sustainable/realizable mechanism. Our numerical experiments and experience show that $S_f\in[0.85,\,0.95]$ works fine; however, a user can use different $S_f$ as per the applications . The formulation requires solving four forward (two for flow and two for structure) problems and three adjoints (one for each objective and one for g$_2$).

\begin{figure*}
	\centering
	\begin{subfigure}[t]{0.40\textwidth}
		\centering
		\includegraphics[scale=0.90]{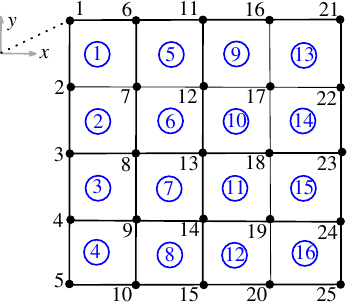}
		\caption{Element and node numbering scheme}
		\label{fig:CoNN}
	\end{subfigure}
	\,
	\begin{subfigure}[t]{0.4\textwidth}
		\centering
		\includegraphics[scale=0.90]{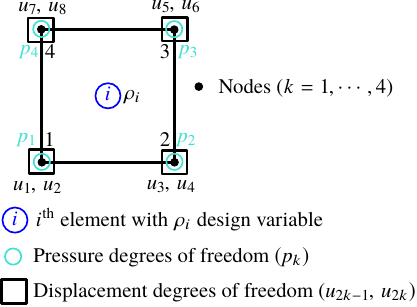}
		\caption{Local format DOFs scheme and nomenclature}
		\label{fig:DOFs}
	\end{subfigure}
	\caption{Figure displays the mesh grid format (element and node numbering scheme) and nomenclature of element~$i$. ({\subref{fig:CoNN}}) $\texttt{nlex} = 4$, $\texttt{nley} = 4$,$\texttt{Lnode} = {1,\,2,\,3,\,4,\,5}$, $\texttt{Tnode} = {1,\,6,\,11,\,16},\,21$, $\texttt{Rnode} = {21,\,22,\,23,\,24,\,25}$, $\texttt{Bnode} = {5,\,10,\,15,\,20,\,25}$; \texttt{nelx} and \texttt{nely} represent the number of elements in $x-$ and $y-$directions, respectively. \texttt{Lnode}, \texttt{Rnode}, \texttt{Tnode}, and \text{Bnode} denote nodes constituting left, right, top, and bottom edges, respectively. ({\subref{fig:DOFs}}) Local format DOFs scheme and nomenclature of element~$i$. Each node is characterized by two displacements and one pressure degree of freedom.} \label{fig:Nomenclature}
\end{figure*} 
\subsection{Sensitivity analysis}\label{Sec:Sensitivity analysis}
This section provides the sensitivity analyses for the objective and constraint functions. The derivatives of these functions are determined using the adjoint-variable method. One can write the augmented performance function $\mathcal{L}_0$ for the objective as
\begin{equation}\label{Eq:augmentedperformance}
	\mathcal{L}_0 = f_0 + \trr{{^1}\bm{\lambda}}_r ( \mathbf{A}_r\mathbf{p}_r) +  \trr{{^2}\bm{\lambda}}_r \left(\mathbf{K}_r\mathbf{u}_r +\mathbf{T} \mathbf{p}_r\right),
\end{equation}
Differentiating Eq.~\ref{Eq:augmentedperformance} with respect to the physical design variable~$\bar{\bm{\rho}}_r$,  one gets
\begin{equation}\label{Eq:Lagrangeder}
	\begin{split}
		\frac{\partial \mathcal{L}_0}{\partial \bar{\bm{\rho}}_r} = &\frac{\partial f_0}{\partial \bar{\bm{\rho}}_r} + 
		\frac{\partial f_0}{\partial\mathbf{u}_r}\frac{\partial\mathbf{u}_r}{\partial \bar{\bm{\rho}}_r}+ {^1}\bm{\lambda}_r^\top \left(\frac{\partial\mathbf{A}_r}{\partial\bar{\bm{\rho}}_r}\mathbf{p}_r\right) \\ &+ {^1}\bm{\lambda}_r^\top\left(\mathbf{A}_r\frac{\partial\mathbf{p}_r}{\partial\bar{\bm{\rho}}_r}\right)+ {^2}\bm{\lambda}_r^\top \left(\frac{\partial\mathbf{K}_r}{\partial \bar{\bm{\rho}}_r}\mathbf{u}_r + \mathbf{K}_r\frac{\partial\mathbf{u}_r}{\partial \bar{\bm{\rho}}_r}\right) \\ & + {^2}\bm{\lambda}_r^\top \left(\frac{\partial\mathbf{T}}{\partial \bar{\bm{\rho}}_r}\mathbf{p}_r + \mathbf{T}\frac{\partial\mathbf{p}_r}{\partial \bar{\bm{\rho}}_r}\right)\\
		=&  \frac{\partial f_0}{\partial \bar{\bm{\rho}}_r}+{^2}\bm{\lambda}_r^\top \left(\frac{\partial\mathbf{K}_r}{\partial \bar{\bm{\rho}}_r}\mathbf{u}_r\right) + {^1}\bm{\lambda}_r^\top \left(\frac{\partial\mathbf{A}_r}{\partial\bar{\bm{\rho}}_r}\mathbf{p}_r\right) \\ &+ \underbrace{\left(\bm{l}^\top + {^2}\bm{\lambda}_r^\top \mathbf{K}_r\right)}_{\mathcal{T}_1}\frac{\partial\mathbf{u}_r}{\partial \bar{\bm{\rho}}_r} + \underbrace{\left({^1}\bm{\lambda}_r^\top\mathbf{A}_r + {^2}\bm{\lambda}_r^\top \mathbf{T} \right)}_{\mathcal{T}_2}\frac{\partial\mathbf{p}_r}{\partial\bar{\bm{\rho}}_r}.
	\end{split}
\end{equation} 
With respect to the  fundamentals of the adjoint-variable method, one chooses ${^1}\bm{\lambda}_r$ and  ${^2}\bm{\lambda}_r$ such that $\mathcal{T}_1 = 0$ and $\mathcal{T}_2 = 0$, which give
\begin{equation} \label{Eq:LagrangeMultiplier}
	\begin{aligned}
		{^2}\bm{\lambda}_r^\top = -\bm{l}^\top\inv{\mathbf{K}}_r,\quad
		{^1}\bm{\lambda}_r^\top = -{^2}\bm{\lambda}_r^\top \mathbf{T} \inv{\mathbf{A}}_r = \bm{l}^\top\inv{\mathbf{K}}_r\mathbf{T} \inv{\mathbf{A}}_r.
	\end{aligned}
\end{equation}
Note that $\frac{\partial f_0}{\partial \bar{\bm{\rho}}_r}=0$, thus, we get 
\begin{equation}\label{Eq:senstivities_Obj}
	\frac{\partial {f_0}}{\partial \bar{\bm{\rho}}_r} = -\bm{l}^\top\inv{\mathbf{K}}_r \frac{\partial\mathbf{K}_r}{\partial \tilde{\bm{\rho}}_r}\mathbf{u}_r  \underbrace{+\bm{l}^\top\inv{\mathbf{K}}_r\mathbf{T} \inv{\mathbf{A}}_r\frac{\partial\mathbf{A}_r}{\partial\bar{\bm{\rho}}_r}\mathbf{p}_r}_{\text{Load sensitivities}}.
\end{equation}
One notes that load sensitivities alter the total objective sensitivity (Eq.~\ref{Eq:senstivities_Obj}). In the case of constant actuating scenarios, only the first term in Eq.~\ref{Eq:senstivities_Obj} appears. Likewise, one finds the derivative of constraint g$_2$ with respect to the physical design vector as
\begin{equation}\label{Eq:senstivities_cons2}
	\frac{\partial {\text{g}_2}}{\partial \bar{\bm{\rho}}_r} = \frac{-\frac{1}{2}\mathbf{u}^\top_r \frac{\partial\mathbf{K}_r}{\partial \bar{\bm{\rho}}_r}\mathbf{u}_r + \mathbf{u}^\top_r \mathbf{T} \mathbf{A}^{-1}_r\frac{\partial\mathbf{A}_r}{\partial\bar{\bm{\rho}}_r}\mathbf{p}_r}{SE^*},
\end{equation}
where $SE^*$ is the permitted strain energy (see Alg.~\ref{Alg:g2}). Determining the derivative of g$_1$ is straightforward~\citep{kumar2022honeytop90}. Finally, we use the chain rule to determine the objective and constraints derivatives with respect to the design variable as
\begin{equation}\label{Eq:objderivative}
	\frac{ \partial{f}}{\partial \bm{\rho}}_r = \frac{\partial {f}}{\partial \bar{\bm{\rho}}_r}\frac{\partial \bar{\bm{\rho}}_r}{\partial \tilde{\bm{\rho}}_r} \frac{\partial \tilde{\bm{\rho}}_r}{\partial \bm{\rho}_r},
\end{equation}
where $f$ represents objective/constraint functions. One uses Eq.~\ref{Eq:senstivities_Obj} and Eq.~\ref{Eq:senstivities_cons2} to determine $\frac{\partial {f}}{\partial \bar{\bm{\rho}}_r}$. $\frac{\partial \bar{\bm{\rho}}}{\partial \tilde{\bm{\rho}}_r}$ and $\frac{\partial \tilde{\bm{\rho}}}{\partial {\bm{\rho}}_r}$ are determined using Eq.~\ref{Eq:der_PHYVariable_filvariable} and Eq.~\ref{Eq:der_filVariable_ovariable}, respectively. Next, we discuss the implementation of the MATLAB code, \texttt{SoRoTop}.
\section{MATLAB implementation}\label{sec:MATLABImple}
\begin{figure*}
	\centering
	\includegraphics[scale=1]{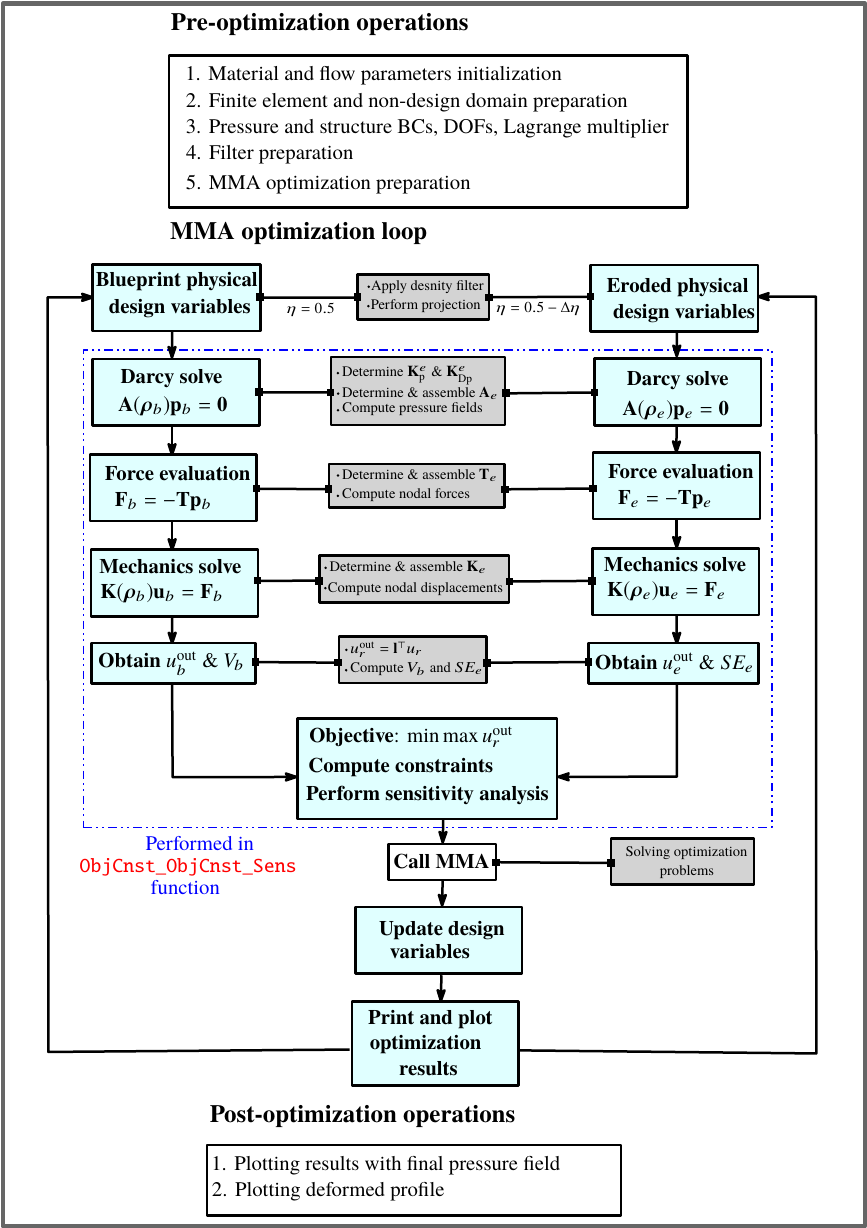}
	\caption{Flowchart for \texttt{SoRoTop} code} \label{fig:flowchart}
\end{figure*}

This section provides a detailed description of the presented code, \texttt{SoRoTop}. The code (Appendix~\ref{sec:sorotop}) and its extensions for different soft mechanisms are provided as supplementary material to the paper. One can call the code from the MATLAB command window as
\begin{lstlisting}[basicstyle=\scriptsize\ttfamily,breaklines=true,numbers=none,frame=tb]
	SoRoTop(nelx,nely,volfrac,sefrac,penal,rmin,kss,etaf,betaf,lst,betamax,delrbst,maxit)
\end{lstlisting}
where \texttt{nelx} and \texttt{nely} are the number of elements in the $x-$ and $y-$directions, respectively, \texttt{volfrac} (Eq.~\ref{Eq:Optimizationequation}) indicates the permitted volume fraction, $\texttt{sefrac}$, a user-defined parameter, controls the stiffness of the mechanisms (Algorithm~\ref{Alg:g2}),  \texttt{penal} (Eq.~\ref{Eq:SIMP}) denotes the penalty parameter of the SIMP scheme, which is set equal to 3, \texttt{kss} represents work-piece stiffness at the output location, and $\texttt{rmin}$ is the filter radius. \texttt{etaf} and \texttt{betaf} are related to the flow coefficient (Eq.~\ref{Eq:Flowcoefficient}) and drainage term (Eq.~\ref{Eq:drainageterm}) and \texttt{lst} decides whether load sensitivities are regarded or not. $\texttt{lst} = 1$ indicates that the load sensitivities are included, whereas $\texttt{lst} = 0$ directs otherwise. $\texttt{betamax}$ is the maximum $\beta$ permitted/used for the Heaviside projection (Eq.~\ref{Eq:projectionFilt}). $\texttt{delrbst}$ represents the $\Delta\eta$  of the robust formulation. $\texttt{maxit}$ is the maximum number of the MMA iteration set for the topology optimization process. Discretization and DOFs (in local format) schemes are depicted in Fig.~\ref{fig:Nomenclature}. The former is in global numbering style (Fig.~\ref{fig:CoNN}), whereas a local numbering scheme is used to demonstrate  DOFs for element~$i$ (Fig.~\ref{fig:DOFs}). Figure~\ref{fig:flowchart} displays the main steps of \texttt{SoRoTop} MATLAB code. Steps performed in function $\xpmt{ObjCnst\_ObjCnst\_Sens}$ are marked inside a blue rectangle (Fig.~\ref{fig:flowchart}).

The code consists of the following subroutines (Fig.~\ref{fig:flowchart}):
\begin{enumerate}
	\item \texttt{Material and flow parameters initialization}
	\item \texttt{Finite element and non-design domain preparation}
	\item \texttt{Pressure and structure BCs, DOFs, Lagrange multipliers}
	\item \texttt{Filter preparation}
	\item \texttt{MMA optimization preparation}
	\item \texttt{MMA optimization loop}
	\item \texttt{Plotting results with final pressure field}
	\item \texttt{Plotting deformed profile}
	\item \texttt{Analyses performing function: $\xpmt{ObjCnst\_ObjCnst\_Sens}$}
\end{enumerate}

Next, we describe each subroutine in detail.
\subsection{\texttt{PART 1: Material and flow parameters initialization (lines~2-7)}}
Lines~2-7 of the code provide material and flow parameters for the design problem. Young's modulus of the material $E_1$ is indicated on line~3 by $\xpmt{E1}$, and that for the void element is mentioned on line~4 by \texttt{Emin}. $\xpmt{E1} =1$   and $\xpmt{Emin} = \xpmt{E1}\times \SI{1e-6}{}$ are set. $\xpmt{nu}$ represents Poisson's ratio set to 0.3 (line~5). The flow coefficient of the void element $K_v$ is indicated by $\xpmt{Kv}$. As indicated in Sec.~\ref{Sec:Design-dependent_PL}, $\xpmt{Kv}=1$ is set on line~6. $\xpmt{epsf}$ denotes the flow contrast  $\epsilon$ (line~6). $\xpmt{epsf} = \SI{1e-7}{}$ is taken, i.e., the flow coefficient of the element with $\rho =1$, $K_s$ equals to $\SI{1e-7}{}$. The parameters $r$ (line~6), $\Delta s$ (line~6) and $Ds$ (line~7) related to the drainage are denoted by $\xpmt{r}$, $\xpmt{Dels}$, and $\xpmt{Ds}$, respectively (Sec.~\ref{Sec:Design-dependent_PL}).  $\xpmt{r} = 0.1$ and $\xpmt{Dels} = 2$ are set (line~6). $\xpmt{Kvs}$ indicates $K_v-K_s = \xpmt{Kv (1-epsf)}$ (line~7). 
\subsection{\texttt{PART 2: Finite element and non-design domain preparation (lines~8-38)}}
Lines~8-38 present finite element, non-design domain, and output DOF preparation. Line~9 determines the total number of elements and nodes participating in the discretized domain by $\xpmt{nel}$ and $\xpmt{nno}$. Lines~10-12 are the standard procedure for determining element DOFs~\citep{andreassen2011efficient,kumar2023TOPress}. $\xpmt{Udofs}$ provides element-wise DOFs, i.e., $i^\text{th}$ row of $\xpmt{Udofs}$ gives DOFs associated with element~$i$ (line~12). Note that $n^\text{th}$ node will have $2n-1$ and $2n$ displacement DOFs in $x-$ and $y-$directions, respectively and $n$ pressure DOF (Fig.~\ref{fig:DOFs}). Nodes constituting left and right edges are determined on line~13 and are indicated via $\xpmt{Lnode}$ and $\xpmt{Rnode}$, respectively. Likewise, on line~14, nodes constituting the bottom and top edges of the domain are found and are denoted via  $\xpmt{Bnode}$ and $\xpmt{Tnode}$, respectively. Line~15 determines different DOFs required for the finite element analysis. Matrices $\xpmt{Pdofs}$  and $\xpmt{allUdofs}$ are used to store DOFs associated with pressure and displacement fields, respectively. $i^\text{th}$ columns of these matrices provide pressure and displacement DOFs of the nodes constituting element~$i$. \texttt{allPdofs} indicates the entire pressure DOFs for the parameterized domain

We determine the element flow matrix due to Darcy law ($\mathbf{Kp}$) on line~16, and that due to drainage term ($\mathbf{KDp}$) on line~17~\citep{kumar2023TOPress}. Matrices~$\mathbf{Kp}$ and $\mathbf{KDp}$ are indicated via $\xpmt{Kp}$ and $\xpmt{KDp}$, respectively, considering unit flow and drainage coefficients. Elemental transformation matrix (Eq.~\ref{Eq:Forcepressureconversion}) is indicated by $\xpmt{Te}$ and is determined on line~18~\citep{kumar2023TOPress}. Element stiffness matrix indicated by $\xpmt{Ke}$ is determined on line~23 per~\citep{andreassen2011efficient} for unit Young's modulus. Matrices required to assemble $\mathbf{KDp}$, $\mathbf{KDp}$, $\xpmt{Te}$ and $\xpmt{Ke}$ to determine corresponding global matrices are determined on lines~24-29. Line~30 and line~32 provide function handles for the projection interpolation and its derivative, respectively. On line~34, a grid of elements is generated and recorded in $\xpmt{elNrs}$. Using $\xpmt{elNrs}$, the non-design solid region is determined on line~35 in vector $\xpmt{s1}$. The elements associated with non-design solid and void regions are determined in matrices $\xpmt{NDS}$ and $\xpmt{NDV}$ on line~36, respectively. Union of $\xpmt{NDS}$ and $\xpmt{NDV}$ gives elements whose design variables do not change with optimization iterations. They are initialized using $\xpmt{deal}$ MATLAB function (line~36). Active design variables are determined and stored in $\xpmt{act}$ array on line~37. The DOF corresponding to the output node is noted in $\xpmt{opdof}$ on line~38. 
\subsection{\texttt{PART 3: Pressure and structure BCs, DOFs, Lagrange multiplier(lines~39-49)}}
Lines~39-49 initialize parameters related to pressure and structure boundary conditions, pressure loads, fixed and free DOFs for pressure and displacement fields, Lagrange multipliers for the adjoint sensitivity calculation, and displacement and pressure vectors.

Vector $\xpmt{PF}$ indicates $\mathbf{p}$ field  (Eq.~\ref{Eq:PDEsolutionpressure}), which is initialized on line~40. Scalar $\xpmt{Pin}$ stores the applied pressure load value (line~40). Line~41 modifies $\xpmt{PF}$  per given pressure loading conditions. Array $\xpmt{fixedPdofs}$  gives nodes corresponding to given pressure conditions (line~43). Free DOFs for the pressure load are determined on line~43 and are recorded in array~$\xpmt{freePdofs}$. Matrix~$\xpmt{pfixeddofsv}$ gives fixed pressure DOFs in the first column and corresponding values in the second column. Fixed displacement DOFs are indicated on line~46 by array $\xpmt{fixedUdofs}$. Line~46 evaluates the free displacement DOFs using $\xpmt{setdiff}$ MATLAB function with inputs vectors $\xpmt{allUdofs}$ and $\xpmt{fixedUdofs}$. Lines~47-48 initialize vectors $\xpmt{L}$, $\xpmt{U}$, $\xpmt{lam1}$, $\xpmt{lam2}$ and $\xpmt{mu2}$. $\xpmt{L}$ represents  vector $l$ (Eq.~\ref{Eq:Optimizationequation}), $\xpmt{U}$ is the global displacement vector $u$ (Eq.~\ref{Eq:Optimizationequation}), $\xpmt{lam1}$ and $\xpmt{lam2}$ are used to determine the objective sensitivity (Eq.~\ref{Eq:LagrangeMultiplier}), whereas for determining $_2$ constraint (eq.~\ref{Eq:Optimizationequation}), Lagrange multiplier $\xpmt{mu2}$ is used in the function $\xpmt{Obj\_ObjSens}$. Vector $\xpmt{L}$ is modified corresponding to the output deformation DOF on line~49.
\subsection{\texttt{PART 4: Filter preparation (lines~50-70)}}
The filter preparation is performed on lines~50-70 per~\citep{andreassen2011efficient}. One can also directly use the $\xpmt{imfilter}$ MATLAB function for filtering operation as done in~\citep{ferrari2021topology,kumar2023TOPress}. $\mathbf{Hs}$ (Eq.~\ref{Eq:HSvalue}) is determined on line~70 and is stored in matrix~$\xpmt{Hs}$. 
\subsection{\texttt{PART 5: MMA optimization preparation (lines~71-83)}}
Lines~71-83 prepare the required parameters for the MMA optimizer. We have used the default setting available in the MMA for the $\min-\max$ (Eq.~\ref{Eq:Optimizationequation}) optimization problems. The design vector $\xpmt{x}$ is initialized on line~72 to zeros. It is updated per the given $\xpmt{volfrac}$, determined $\xpmt{NDS}$ and $\xpmt{NDV}$ on line~73 for the active set of design variables. Parameters $\xpmt{nMMA}$ (line~74) and $\xpmt{mMMA}$ (line~75) are used to indicate the number of design variables and constraints, respectively, that MMA is going to handle for the given problem. $\xpmt{pMMA}$ indicates the number of objectives considered within the robust optimization formulation. Herein, we consider output deformation values of the blueprint and eroded designs, i.e., $\xpmt{pMMA} = 2$  is set on line~74. Likewise, $\xpmt{qMMA}$ provides the number of constraints applied. A volume and a strain energy constraints are applied on the blueprint and eroded designs, respectively, i.e., $\xpmt{qMMA} = 2$ (line~74). The active design variables will be used in the MMA, which are copied in vector $\xpmt{xMMA}$ on line~73. The filtered design vector is initialized on line~75 in $\xpmt{xTilde}$. $\xpmt{mvLt}$ indicates the external move limit of the MMA, which is set to 0.1 (line~75). On line~76, minimum and maximum values of the design vector are initialized in vectors $\xpmt{xminvec}$ and $\xpmt{xmaxvec}$, respectively. The lower and upper limits on the design vector are set in vectors $\xpmt{low}$ and $\xpmt{upp}$, respectively, on line~77. The parameters $\xpmt{cMMA}$, $\xpmt{dMMA}$ and $\xpmt{a0}$  of the MMA~\citep{svanberg1987} are initialized on line~78. On line~79, parameter $\xpmt{aMMA}$ is defined. Optimization is an iterative process; the old design vectors one and two iterations ago are recorded in vectors $\xpmt{xold1}$ and $\xpmt{xold2}$, respectively. These vectors are initialized on line~80. A scalar $\xpmt{loop}$ counts the MMA iterations. The MMA loop is repeated until a convergence parameter $\xpmt{change}$ is lower than a  $\SI{1e-4}{}$ or provided number of iterations (\xptt{maxit}) is reached. The sharpness parameter $\beta$ of the projection function (Eq.~\ref{Eq:projectionFilt}) is denoted by $\xpmt{betap}$. $\xpmt{betap}$, $\xpmt{loop}$ and $\xpmt{change}$ are initialized on line~81 using $\xpmt{deal}$ MATLAB function.  On line~82, steps $\eta$ (Eq.~\ref{Eq:projectionFilt}) for the blueprint ($\xpmt{etai}$) and eroded ($\xpmt{etae}$) designs are set to 0.5 and $0.5 + \xpmt{delrbst}$. A constant $\xpmt{constadd} = 10000$ is set and used on line~107 so that $\xpmt{fval}$ of the MMA shall not become negative during optimization.

\subsection{\texttt{PART 6: MMA optimization loop (lines~84-120)}}\label{Sec:MMA optimizationloop}
Optimization iteration is performed on lines~84-120 using the parameters and variables defined above. Optimization starts with \texttt{while} loop on line~85. The \texttt{while} loop gets terminated if $\xpmt{loop}$ equals the maximum number of permitted optimization iterations $\xpmt{maxit}$ and  $\xpmt{change}$ gets a value lower than $\SI{1e-4}{}$. The optimization procedure is divided into eight major subparts detailed below.

\subsubsection{\texttt{PART 6.1: Compute blueprint and eroded physical design variables (lines~87-90)}}
Active filtered design variables $\xpmt{xTilde (act)}$ are assigned per $\xpmt{xMMA}$ on line~88, which is filtered using $\xpmt{Hs}$ (evaluated on line~70). Further, using vectors $\xpmt{NDS}$ and $\xpmt{NDV}$, the filter vector $\xpmt{xTilde}$ is updated. Physical design vectors for the blueprint and eroded designs are represented via $\xpmt{xphysb}$ and $\xpmt{xphyse}$, respectively. They are determined using the function handle $\xpmt{IFprj}$ with inputs $\left\{\xpmt{xTilde},\,\xpmt{etab},\,\xpmt{betap}\right\}$ and $\left\{\xpmt{xTilde},\,\xpmt{etae},\,\xpmt{betap}\right\}$ respectively. They are updated using vectors $\xpmt{NDS}$ and $\xpmt{NDV}$.
\subsubsection{\texttt{PART 6.2: Performing blueprint design analysis using ObjCnst\_ObjCnst\_Sens function (lines~91-93)}}
$\xpmt{ObjCnst\_ObjCnst\_Sens}$ function is called with $\xpmt{xphysb}$ as one of the input variables to perform analysis for the blueprint design. We record the required output of $\xpmt{Obj\_ObjSens}$. The output deformation and volume of the blueprint design are recorded in $\xpmt{objb}$ and $\xpmt{volb}$ (line~92). Blueprint objective and volume constraint sensitivities are noted in $\xpmt{objsensb}$ and $\xpmt{volsensb}$, respectively. We also note the blueprint design displacement field, nodal forces, and pressure field in $\xpmt{Ub}$, $\xpmt{Fb}$, and $\xpmt{PFb}$, respectively.
\subsubsection{\texttt{PART 6.3: Performing eroded design analysis using ObjCnst\_ObjCnst\_Sens function (lines~94-96)}}
Similar to the blueprint design analysis, we call  $\xpmt{ObjCnst\_ObjCnst\_Sens}$ function for eroded design but with $\xpmt{xphyse}$ as one of the physical input variables. The objective and strain energy are recorded in $\xpmt{obje}$ and $\xpmt{SEe}$, and their sensitivities are noted in $\xpmt{objsense}$ and $\xpmt{SEsense}$, respectively.
\subsubsection{\texttt{PART 6.4: Filtering and projecting sensitivities of objectives and constraints  (lines~97-101)}}
Sensitivities of obtained objective ($\xpmt{objsensb},\,\xpmt{objsense}$) and constraints ($\xpmt{volsensb},\,\xpmt{SEsense}$) are filtered and projected using $\xpmt{Hs}$ and  $\xpmt{dIFprj}$. $\left\{\xpmt{xphysb},\,\xpmt{etab},\,\xpmt{betap}\right\}$ and $\left\{\xpmt{xphyse},\,\xpmt{etae},\,\xpmt{betap}\right\}$ are respectively used as input variables to $\xpmt{dIFprj}$ for the blueprint and eroded designs. 
\subsubsection{\texttt{PART 6.5: Stacking constraints and their sensitivities (lines~102-105)}}
The current volume of the blueprint design and strain energy of the eroded design are stacked in $\xpmt{constr}$ on line~103. $\xpmt{constrsens}$ is used to stack sensitivities of the volume and strain energy constraints on line~104. These are performed to suit the requirements of the MMA. $\xpmt{normf}$ (line~105) normalizes objectives and sensitivities.       
\subsubsection{\texttt{PART 6.6: Setting and calling MMA optimizer (lines~106-111)}}
This section calls the $\xptt{mmasub}$ function of the MMA optimizer on line~111 to solve the $\min-\max$ optimization problem (Eq.~\ref{Eq:Optimizationequation}). The default setting is used, wherein objectives and constraints are treated as constraints~\citep{svanberg2007mma}. The reader can refer to \cite{svanberg2007mma} for a detailed description of implementing $\min-\max$ optimization problem using the MMA~\citep{svanberg1987}. $\xpmt{fval}$ is determined on line~107, wherein $\xpmt{constadd}$ is used to avoid getting negative entries for $\xpmt{fval}$ during optimization. $\xpmt{dfdx}$ is determined on line~108 using the objectives' and constraints' derivatives. On line~109, $\xpmt{xminvec}$ and $\xpmt{xmaxvec}$ are updated using the move limit.
\subsubsection{\texttt{PART 6.7: Updating MMA/design variables and projection parameter $\beta$ (lines~112-116)}}
This part of the code updates the design variable vector $\xpmt{xMMA}$ using the solution obtained from the MMA on line~115. The vectors $\xpmt{xold1}$ and $\xpmt{xold2}$ are updated on line~113. The new solution of the MMA is recorded in $\xpmt{xnew}$ on line~113. $\xpmt{change}$ is determined on line~114 using the new and previous design variables. $\beta$ ($\xpmt{betap}$) of the projection filter (Eq.~\ref{Eq:projectionFilt}) is updated on line~116 at every 25th MMA iterations until it reaches the assigned maximum value $\xpmt{betamax}$. 
\subsubsection{\texttt{PART 6.8: Print and plot the results (lines~117-120)}}
This part of the code print some important values ($\xpmt{loop}$, $\xpmt{fval(1)}$, i.e., $\xpmt{objb}$, $\xpmt{fval(2)}$, i.e., $\xpmt{obje}$, $\xpmt{volb}$ using $\xpmt{mean(xphysb)}$, and $\xpmt{change}$ of the optimization. Finally,  the optimization evolution is plotted on line~119 using $\xpmt{colormap}$ and $\xpmt{imagesc}$ MATLAB functions.  Next we describe post-optimization operations and $\xpmt{ObjCnst\_ObjCnst\_Sens}$ function.

\subsection{Post-optimization operations}\label{sec:post-optimization}
We perform two operations after optimization ends. Firstly, the optimized result is plotted with the final pressure field $\xpmt{PE}$, described on lines~121-138 (PART~7). Nodal information is recorded on line~123 in matrix $\xpmt{node}$. Each row of $\xpmt{node}$ has three entries, i.e., $i^\text{th}$ row indicates node~$i$ (first entry) with its $x-$ (2nd entry) and $y-$ (third entry) coordinates. Likewise, matrix $\xpmt{elem}$ contains element information with element number and its nodes. Matrices $\xpmt{X}$ (line~125) and $\xpmt{Y}$ (line~125) contain the $x-$ and $y-$ coordinates of nodes in element-wise sense. Matrix $\xpmt{Y1}$ is defined for plotting symmetric results. On line~127, the final pressure field (nodal pressure) is converted to elemental pressure loads and are stored in vector $\xpmt{elemP}$. Lines~128-129 are used to plot the full optimized results using $\xpmt{patch}$ MATLAB function, wherein $\xpmt{X}$, $\xpmt{Y}$, $\xpmt{Y1}$ and $\xpmt{xphysb}$ quantities are used. Lines~130-138 plot the final pressure field using $\xpmt{patch}$ MATLAB function.

Secondly, the scaled-deformed profile is plotted on lines~139-155 (PART~8).  The deformed nodal coordinates are stored in $\xpmt{xn}$ using $\xpmt{node}$ and $\xpmt{Ub}$ information (lines~141-144). $\xpmt{Xn}$ and $\xpmt{Yn}$ are determined using $\xpmt{xn}$, they stored the $x-$ and $y-$nodal coordinates in matrix form. Matrix $\xpmt{Yn1}$ (line~144) provides $y-$coordinates for symmetry half design. Finally, lines~147-155 plots the deformed profile using $\xpmt{patch}$ MATLAB function. 

\subsection{PART 9: $\xpmt{ObjCnst\_ObjCnst\_Sens}$ function}
This function is written to perform FEM analyses pertaining to pressure field, objective and constraints evaluation, and derivative of objective and constraints determination. The function is called twice in the MMA optimization loop (Sec.~\ref{Sec:MMA optimizationloop}) for the blueprint and eroded designs analyses. The function is provided with the code on lines~156-199 as
\begin{lstlisting}[basicstyle=\scriptsize\ttfamily,breaklines=true,numbers=none,frame=tb]
	function[obj,objsens,vol,volsens, SEc, SEsens,U,F,PF] = ObjCnst_ObjCnst_Sens(xphys,nel,E1,Emin,penal,Kv,kvs,epsf,Ds,etaf,betaf,Udofs,freeUdofs,Pdofs,pfixeddofsv,fixedPdofs,freePdofs,iP,jP,iT,jT,iK,jK,Kp,KDp,Te,ke,outputdof,kss,loop,IFprj,dIFprj,L,U,lam1,lam2,mu2,lst,volfrac,sefrac)
\end{lstlisting}
whose input and output variables are  explained above (in subroutines 1-5 of $\xpmt{SoRoTop}$ code). We provide a detailed description  function below that contains six subparts. 

The first (lines~159-167) subpart provides the pressure filed $\xpmt{PF}$, i.e., solves Darcy law (Eq.~\ref{Eq:stateequation}). The flow coefficient $\mathbf{K}$ (Eq.~\ref{Eq:Flowcoefficient}) is indicated by $\xpmt{Kc}$ on line~160, which is determined using $\xpmt{Kv}$ (line~6), $\xpmt{epsf}$ (line~6) and $\xpmt{IFprj}$ (line~30). The drainage coefficient $\mathbf{D}$ (Eq.~\ref{Eq:drainageterm}) is denoted by $\xpmt{Dc}$ on line~161. $\xpmt{Dc}$ is determined using $\xpmt{Ds}$~(line~7) and $\xpmt{IFprj}$~(line~30). The elemental flow matrices of all elements are recorded as a vector form in $\xpmt{Ae}$ on line~162. $\xpmt{Ae}$ is evaluated using $\xpmt{reshape}$ MATLAB function with $\xpmt{Kp}$ (line~16), $\xpmt{KDp}$ (line~17), $\xpmt{Kc}$~(line~160), $\xpmt{Dc}$~(line~161) and $\xpmt{nel}$. Using $\xpmt{Ae}$, $\xpmt{iP}$~(line~24) and $\xpmt{jP}$~(line~25) the global flow matrix $\mathbf{A}$ (Eq.~\ref{Eq:PDEsolutionpressure}) for the Darcy law with drainage term is evaluated on line~163 and recorded in $\xpmt{AG}$. Using the free DOFs of pressure field, i.e., $\xpmt{freePdofs}$ (line~43), the corresponding flow matrix is determined on line~164 and recorded in $\xpmt{Aff}$. We decompose $\xpmt{Aff}$ on line~165 using \linebreak$\xpmt{decomposition}$ MATLAB function and recorded in $\xpmt{dAff\_ldl}$. The pressure field $\mathbf{P}$ (Eq.~\ref{Eq:PDEsolutionpressure}) is determined on line~166 as per Eq.~\ref{Eq:partEq6} and stored in $\xpmt{PF}$. Finally, $\xpmt{PF}$ is updated on line~167 using the given pressure load stored in $\xpmt{pfixeddofsv}$ (line~44).

The second (lines~168-177) subpart gives consistent global nodal load $\mathbf{F}$~(Eq.~\ref{Eq:Forcepressureconversion}) and displacement~(Eq.~\ref{Eq:Optimizationequation}) vectors. On line~169, the element transformation matrix $\mathbf{T}_e$ (Eq.~\ref{Eq:Forcepressureconversion}) is written in a vector form in $\xpmt{Ts}$. Using the matrices $\xpmt{iT}$ (line~26) and $\xpmt{jT}$ (line~27), we find the global transformation matrix $\mathbf{T}$ (Eq.~\ref{Eq:nodalforce}) on line~170 and record in $\xpmt{TG}$. $\mathbf{F}$ (Eq.~\ref{Eq:nodalforce}) is determined on line~171 using matrix $\xpmt{TG}$ and vector $\xpmt{PF}$. Young's modulus interpolation is performed on line~172 in vector $\xpmt{E}$. The $i^\text{th}$ entry of  $\xpmt{E}$ provides Young's modulus of element~$i$. The element stiffness matrices in a vector form using $\xpmt{Ke}$, $\xpmt{E}$ and $\xpmt{nel}$ are arranged in $\xpmt{kss}$ on line~173. The global stiffness matrix $\mathbf{K}$ is evaluated on line~174 and is stored in $\xpmt{KG}$. Matrix $\xpmt{KG}$ is updated corresponding to the output degree of freedom $\xpmt{opdof}$ using the workpiece stiffness $\xpmt{kss}$. On line~176, the global stiffness matrix is decomposed using $\xpmt{decomposition}$ MATLAB function with the Cholesky scheme. $\xpmt{U}$ is determined on line~177. In the third part (lines~178-179), the objective of the mechanism is determined using vectors $\xpmt{L}$ and $\xpmt{U}$.

The fourth (lines~180-187) subpar provides the objective sensitivities per Sec.~\ref{Sec:Sensitivity analysis}. $\xpmt{lam1}$  and $\xpmt{lam2}$  indicate $\lambda_1$ (Eq.~\ref{Eq:LagrangeMultiplier}) and $\lambda_2$ (Eq.~\ref{Eq:LagrangeMultiplier}), respectively. They are determined on line~181 and line~182, respectively. $\xpmt{objsT1}$ determines the first part of the objective sensitivities (Eq.~\ref{Eq:objderivative}). Lines~184-186 evaluate the second part (Eq.~\ref{Eq:objderivative}), i.e., the load sensitivities. $\xpmt{dC1k}$ (line~184) and $\xpmt{dC1d}$ (line~185) determine the Darcy law and drainage parts, respectively, of the load sensitivities. On line~186, vector $\xpmt{objsT2}$ records the load sensitivity terms. $\xpmt{objsens}$ stores the objective sensitivities on line~187. The fifth part determines the volume constraint $\xpmt{vol}$ (line~189) and its sensitivities $\xpmt{volsens}$ with respect to the design variables on line~190.

The last (line~191-199) subpart determines the strain energy sensitivities per Sec.~\ref{Sec:Sensitivity analysis}. On line~192, the permitted strain energy $\xpmt{SE\_perm}$ is determined using $\xpmt{sefrac}$ and strain energy determined at $\xpmt{loop}=1$. $\xpmt{SE\_perm}$ is also saved and loaded on line~192. The strain energy constraint $\xpmt{SEc}$ (Eq.~\ref{Eq:senstivities_cons2}) and its sensitivities are  determined on line~193 and line~199, respectively. $\xpmt{SET1}$ records the first part of the  constraint sensitivities (Eq.~\ref{Eq:senstivities_cons2}), whereas $\xpmt{SET2}$ (Eq.~\ref{Eq:senstivities_cons2}) stores the second part of the sensitivities. $\xpmt{mu1}$ is the Lagrange multiplier needed to determine $\xpmt{SET2}$.  One line~196 and line~197, $\xpmt{dSEk}$, the flow part of the sensitivities, and $\xpmt{dSEd}$, the drainage part of the sensitivities, are determined. Finally, the strain energy sensitivities are determined on line~199 and stored in vector~$\xpmt{SEsens}$. Next, we present numerical results and discussions. 
\begin{figure*}[h!]
	\centering
	\begin{subfigure}{0.45\textwidth}
		\centering
		\includegraphics[scale=0.5]{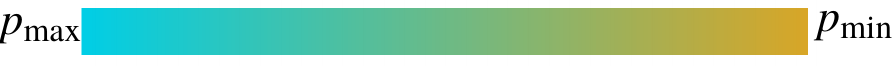}
		\caption{}
		\label{fig:pressurecolorbar}
	\end{subfigure}
	\hfill
	\begin{subfigure}{0.45\textwidth}
		\centering
		\includegraphics[scale=0.5]{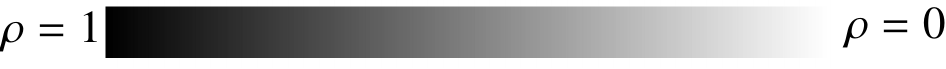}
		\caption{}
		\label{fig:materialcolorbar}
	\end{subfigure}
	\caption{Colorbar schemes for the pressure and material fields are shown in (\subref{fig:pressurecolorbar}) and (\subref{fig:materialcolorbar}), respectively. $P_\text{max} = 1$ bar and $p_\text{min} =0$ bar.} \label{fig:Press_mat_colorbar}
\end{figure*}

\begin{figure}
	\centering
	\includegraphics[scale=1.5]{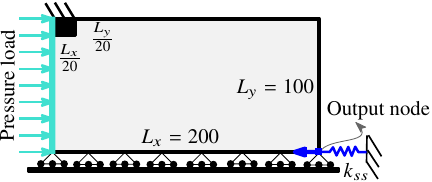}
	\caption{Inverter design domain} \label{fig:InverterDesisn}
\end{figure}

\begin{figure*}[h!]
	\centering
	\begin{subfigure}{0.22\textwidth}
		\centering
		\includegraphics[scale=0.3500]{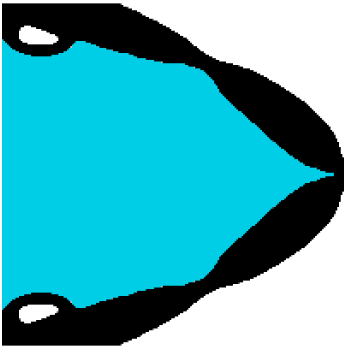}
		\caption{$u^\text{out} = -148.248$}
		\vspace*{-0.5em}
		$\xpmt{lst} =1$
		\label{fig:IN_undef_lst_1}
	\end{subfigure}
	\,
	\begin{subfigure}{0.22\textwidth}
		\centering
		\includegraphics[scale=0.350]{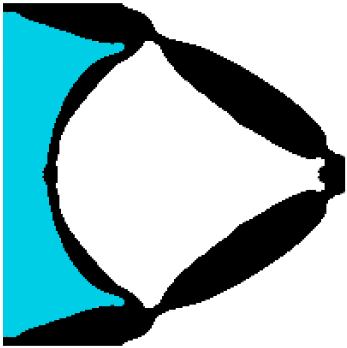}
		\caption{$u^\text{out} =-144.586$}
		\vspace*{-0.5em}
		$\xpmt{lst} =0$
		\label{fig:IN_undef_lst_0}
	\end{subfigure}
	\,
	\begin{subfigure}{0.22\textwidth}
		\centering
		\includegraphics[scale=0.350]{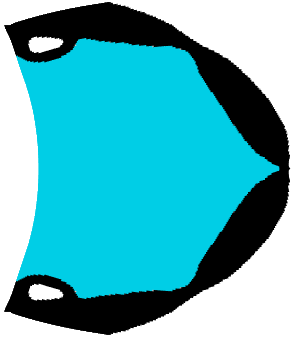}
		\caption{$\xpmt{lst} =1$}
		\label{fig:IN_def_lst_1}
	\end{subfigure}
	\,
	\begin{subfigure}{0.22\textwidth}
		\centering
		\includegraphics[scale=0.350]{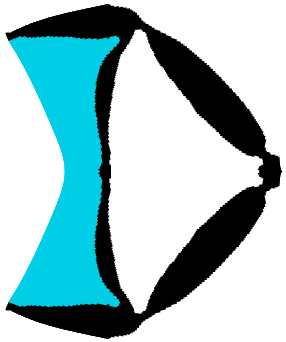}
		\caption{$\xpmt{lst} =0$}
		\label{fig:IN_def_lst_0}
	\end{subfigure}
	\caption{Pneumatically actuated optimized soft inverter mechanisms  with (\subref{fig:IN_undef_lst_1}) $\texttt{lst} =1$ and  (\subref{fig:IN_undef_lst_0}) $\texttt{lst} =0$. Pneumatically actuated optimized soft inverter mechanisms  in deformed configurations with (\subref{fig:IN_def_lst_1}) $\texttt{lst} =1$ and  (\subref{fig:IN_def_lst_0}) $\texttt{lst} =0$.} \label{fig:Inverter_Lst_1}
\end{figure*} 

\begin{figure*}[h!]
	\centering
	\begin{subfigure}{0.23\textwidth}
		\centering
		\includegraphics[scale=0.35]{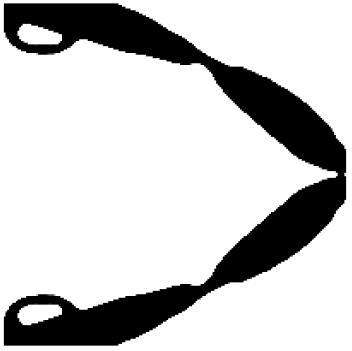}
		\caption{Eroded design}
		\vspace*{-0.1em}
	    $\mathtt{lst=1},\,V_f = 0.23$
		\label{fig:inverter_eroded}
	\end{subfigure}
	\hfill
	\begin{subfigure}{0.23\textwidth}
		\centering
		\includegraphics[scale=0.35]{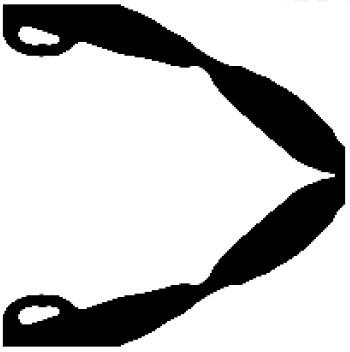}
		\caption{Blueprint design}
		\vspace*{-0.1em}
	$\mathtt{lst=1},\,V_f = 0.25$
		\label{fig:inverter_blueprint}
	\end{subfigure}
	\begin{subfigure}{0.23\textwidth}
		\centering
		\includegraphics[scale=0.35]{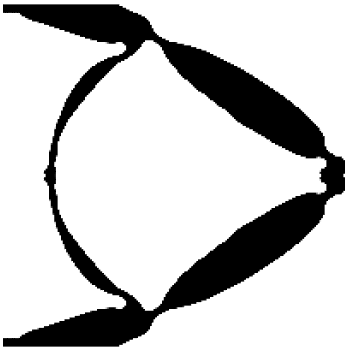}
		\caption{Eroded design}
		\vspace*{-0.1em}
	$\mathtt{lst=0},\,V_f = 0.226$
		\label{fig:inverter_eroded_lst_0}
	\end{subfigure}
	\hfill
	\begin{subfigure}{0.23\textwidth}
		\centering
		\includegraphics[scale=0.35]{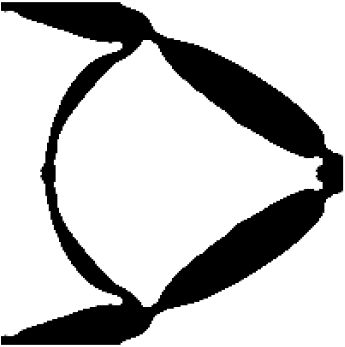}
		\caption{Blueprint design}
		\vspace*{-0.1em}
	$\mathtt{lst=0},\,V_f = 0.25$
		\label{fig:inverter_blueprint_lst_0}
	\end{subfigure}
	\caption{The eroded and blueprint (intermediate) optimized designs for the soft pneumatic inverters obtained with $\mathtt{lst} =1$ and $\mathtt{lst} =0$. $V_f$ indicates the final volume fraction. } \label{fig:inverter_eroded_blueprint}
\end{figure*}

\begin{figure*}[h!]
	\centering
	\begin{subfigure}{0.13\textwidth}
		\centering
		\includegraphics[scale=0.22]{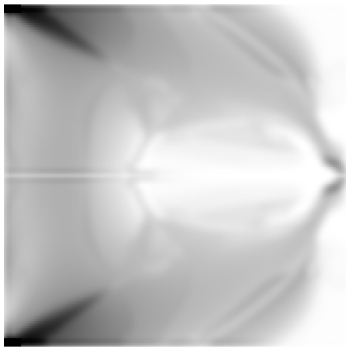}
		\caption{$\mathtt{it=10}$}
		\label{fig:IV_10_lst_1}
	\end{subfigure}
	\hfill
	\begin{subfigure}{0.13\textwidth}
		\centering
		\includegraphics[scale=0.22]{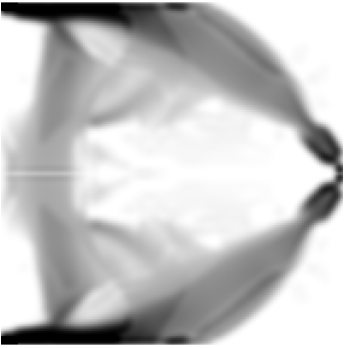}
		\caption{$\mathtt{it=20}$}
		\label{fig:IV_20_lst_1}
	\end{subfigure}
	\hfill
	\begin{subfigure}{0.13\textwidth}
		\centering
		\includegraphics[scale=0.22]{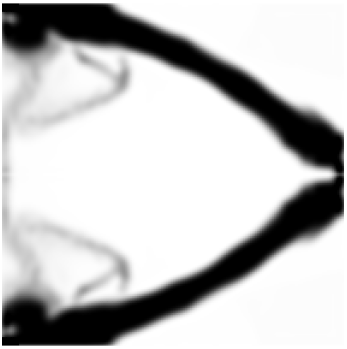}
		\caption{$\mathtt{it=40}$}
		\label{fig:IV_40_lst_1}
	\end{subfigure}
	\hfill
	\begin{subfigure}{0.13\textwidth}
		\centering
		\includegraphics[scale=0.22]{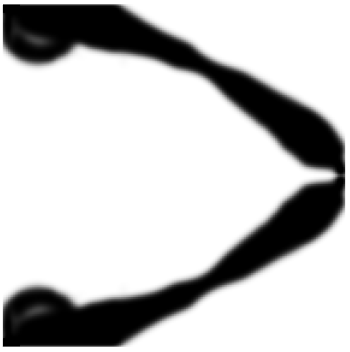}
		\caption{$\mathtt{it=51}$}
		\label{fig:IV_50_lst_1}
	\end{subfigure}
	\hfill
	\begin{subfigure}{0.13\textwidth}
		\centering
		\includegraphics[scale=0.22]{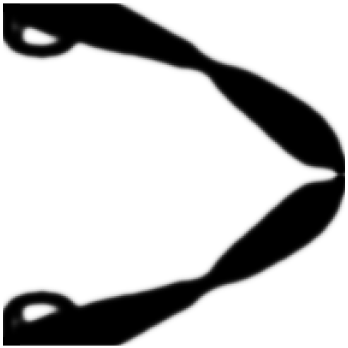}
		\caption{$\mathtt{it=101}$}
		\label{fig:IV_101_lst_1}
	\end{subfigure}
	\hfill
	\begin{subfigure}{0.13\textwidth}
		\centering
		\includegraphics[scale=0.22]{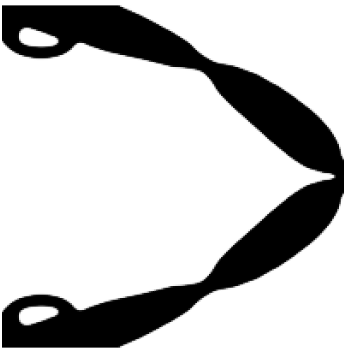}
		\caption{$\mathtt{it=201}$}
		\label{fig:IV_201_lst_1}
	\end{subfigure}
	\hfill
	\begin{subfigure}{0.13\textwidth}
		\centering
		\includegraphics[scale=0.22]{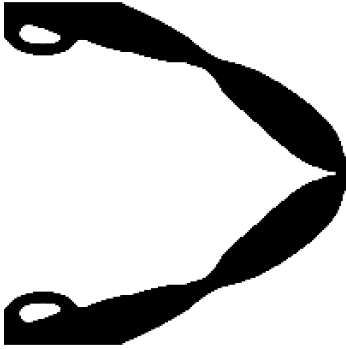}
		\caption{$\mathtt{it=301}$}
		\label{fig:IV_301_lst_1}
	\end{subfigure}
	\hfill
	\begin{subfigure}{0.13\textwidth}
		\centering
		\includegraphics[scale=0.22]{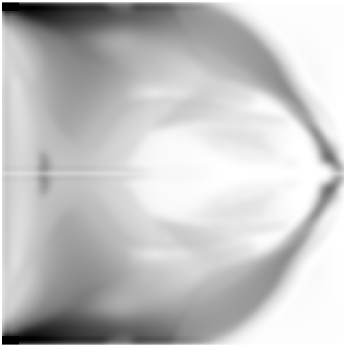}
		\caption{$\mathtt{it=10}$}
		\label{fig:IV_10_lst_0}
	\end{subfigure}
	\hfill
	\begin{subfigure}{0.13\textwidth}
		\centering
		\includegraphics[scale=0.22]{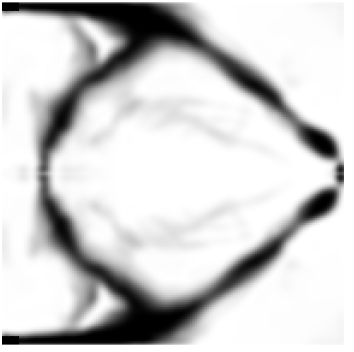}
		\caption{$\mathtt{it=20}$}
		\label{fig:IV_20_lst_0}
	\end{subfigure}
	\hfill
	\begin{subfigure}{0.13\textwidth}
		\centering
		\includegraphics[scale=0.22]{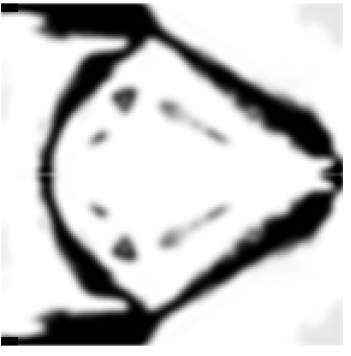}
		\caption{$\mathtt{it=40}$}
		\label{fig:IV_40_lst_0}
	\end{subfigure}
	\hfill
	\begin{subfigure}{0.13\textwidth}
		\centering
		\includegraphics[scale=0.22]{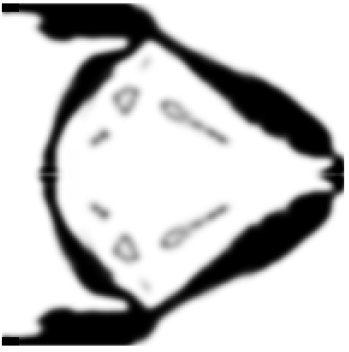}
		\caption{$\mathtt{it=51}$}
		\label{fig:IV_50_lst_0}
	\end{subfigure}
	\hfill
	\begin{subfigure}{0.13\textwidth}
		\centering
		\includegraphics[scale=0.22]{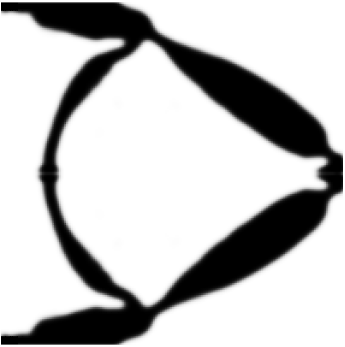}
		\caption{$\mathtt{it=101}$}
		\label{fig:IV_101_lst_0}
	\end{subfigure}
	\hfill
	\begin{subfigure}{0.13\textwidth}
		\centering
		\includegraphics[scale=0.22]{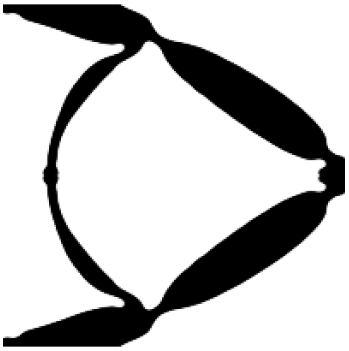}
		\caption{$\mathtt{it=201}$}
		\label{fig:IV_201_lst_0}
	\end{subfigure}
	\hfill
	\begin{subfigure}{0.13\textwidth}
		\centering
		\includegraphics[scale=0.22]{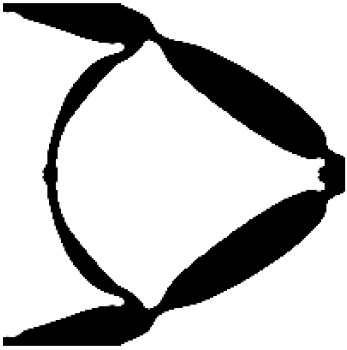}
		\caption{$\mathtt{it=301}$}
		\label{fig:IV_301_lst_0}
	\end{subfigure}
	\caption{Intermediate results for the soft pneumatic inverter. (\subref{fig:IV_10_lst_1}) - (\subref{fig:IV_301_lst_1}) : Obtained with $\mathtt{lst =1}$. (\subref{fig:IV_10_lst_0}) - (\subref{fig:IV_301_lst_0}) : Obtained with $\mathtt{lst =0}$. $\mathtt{it}$ indicates the optimization iteration number. }\label{fig:IV_intermediate}
\end{figure*}
\section{Results and discussions}\label{sec:ResultsDiscussion}
This section provides optimized pneumatically activated soft mechanisms using the presented code and extensions. The employed colorbar schemes for depicting the optimized material and pressure fields are show in Fig.~\ref{fig:Press_mat_colorbar}. 

\subsection{Soft pneumatic inverter}
Herein, soft pneumatic inverters are designed. Such mechanisms in 2D are reported in~\cite{de2020topology,kumar2020topology,kumar2022topological,lu2021topology,kumar2022improved,moscatelli2023pure}, whereas \citet{kumar2021topology} also  report a 3D pneumatic-actuated inverter mechanism. The provided code, $\xpmt{SoRoTop}$, has the default setting for designing a soft inverter mechanism. The half symmetric design domain $L_x\times L_y = 200\times 100$ for the soft pneumatic inverter and its pressure and displacement boundary conditions are depicted in Fig.~\ref{fig:InverterDesisn}. The pressure/pneumatic load is applied on the left edge toward the $+x$ direction, and the output node depicted in the figure is desired to move toward the $-x$ direction. The top corner of the domain is fixed. A non-design solid domain of dimension $\frac{L_x}{20} \times \frac{L_y}{20}$  exists as shown the figure. The symmetric boundary condition is also depicted (Fig.~\ref{fig:InverterDesisn}). One calls $\xpmt{SoRoTop}$ in the MATLAB command windows as
\begin{lstlisting}[basicstyle=\scriptsize\ttfamily,breaklines=true,numbers=none,frame=tb]
	SoRoTop(200,100,0.25,0.95,3,4.8,1,0.1,8,1,128,0.15,400)
\end{lstlisting}
with $\xpmt{nelx} = 200,\,\xpmt{nely} =100,\,\xpmt{volfrac} = 0.25,\,\xpmt{sefrac} = 0.95,\,\xpmt{penal} =3,\,\xpmt{rmin} = 4.8,\,\xpmt{kss}=1;\xpmt{etaf} =0.1,\,\xpmt{betaf}=8,\,\xpmt{lst}=1,\,\xpmt{betamax}=128,\,\xpmt{delrbst}=0.15$ and $\xpmt{maxit}=400$. As $\xpmt{lst}=1$, i.e., the load sensitivities are included in the optimization. 

The obtained symmetric result is transferred into a full design and is displayed in Fig.~\ref{fig:IN_undef_lst_1}. The corresponding deformed profile (scaled) is depicted in Fig.~\ref{fig:IN_def_lst_1}. $u^\text{out} = -148.248$ is obtained, i.e., the determined output deformation of the soft pneumatic inverter is in the desired direction. 

\begin{figure}[ht!]
	\centering
	\begin{tikzpicture}
		\pgfplotsset{compat = 1.3}
		\begin{axis}[ blue,
			width = 0.5\textwidth,
			xlabel=MMA iteration,
			axis y line* = left,
			ylabel= Compliance,
			ymajorgrids=true,
			xmajorgrids=true,
			grid style=dashed ]
			\pgfplotstableread{R1Obj.txt}\mydata;
			\addplot[smooth,blue,mark = *,mark size=1pt,style={thick}]
			table {\mydata};\label{plot1}
		\end{axis}
		\begin{axis}[
			width = 0.5\textwidth,
			axis y line* = right,
			ylabel= Volume fraction,
			axis x line*= top,
			xlabel=MMA iteration,
			ytick = {0.21,0.22,0.23,0.24,0.25},
			yticklabel style={/pgf/number format/.cd,fixed,precision=3},
			legend style={at={(0.95,0.5)},anchor=east}]
			\addlegendimage{/pgfplots/refstyle=plot1}\addlegendentry{Objective}
			\pgfplotstableread{R1vol.txt}\mydata;
			\addplot[smooth,black,mark = square,mark size=1pt,style={thick}]
			table {\mydata};
			\addlegendentry{volume fraction}
		\end{axis}
	\end{tikzpicture}
	\caption{Objective and volume fraction convergence plots for the blueprint soft pneumatic inverter}
	\label{fig:Convergenceplot}
\end{figure}
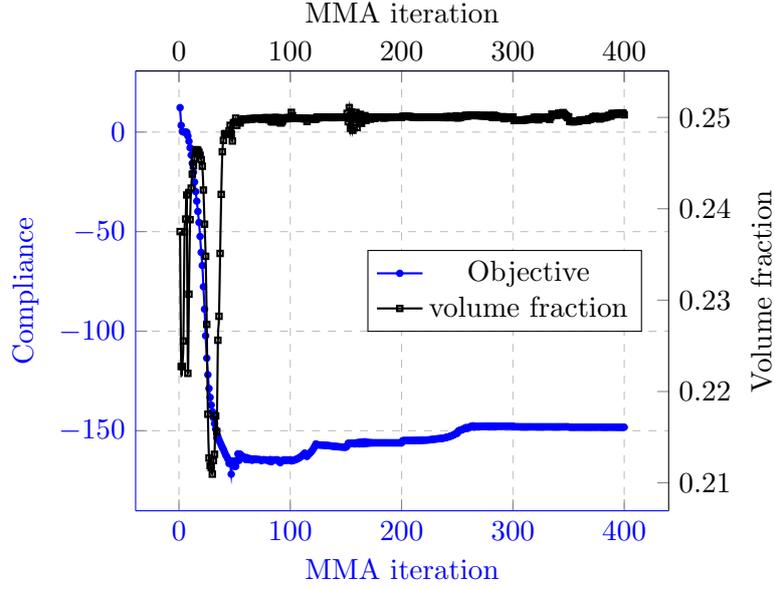

We now use $\xpmt{lst}=0$ while keeping all the input variables of $\xpmt{SoRoTop}$ as above to demonstrate the effects of the load sensitivities on the optimized soft pneumatic inverter. With $\xpmt{lst =0}$, the optimized design and corresponding deformed profiles are shown in Fig.\ref{fig:IN_undef_lst_0} and Fig.~\ref{fig:IN_def_lst_0} , respectively. $u^\text{out} = -144.586$ is obtained in the desired direction. The topology of the optimized design obtained with $\xpmt{lst=0}$ (Fig.~\ref{fig:IN_undef_lst_0}) is different than achieved with $\xpmt{lst=1}$ (Fig.~\ref{fig:IN_undef_lst_1}). The latter has a relatively bigger pressurized region than that of the former. Considering load sensitivities is physically right, it affects the topology of the optimized designs. The current observation aligns with that noticed in~\cite{kumar2023TOPress}  with $\xpmt{TOPress}$ MATLAB code for optimizing pressure loadbearing structures. The deformed profiles (scaled) are depicted in Fig.~\ref{fig:IN_def_lst_1} and Fig.~\ref{fig:IN_def_lst_0} with $\xpmt{lst}=1$ and $\xpmt{lst}=0$, respectively. The obtained output displacements of the optimized mechanisms are in the desired direction, i.e., in the negative $x-$direction. Fig.~\ref{fig:inverter_eroded_blueprint} depicts the eroded and blueprint soft pneumatic inverter optimized designs. The eroded designs contain relatively thinner members than the blueprint designs (Fig.~\ref{fig:inverter_eroded_blueprint}). Fig.~\ref{fig:IV_intermediate} shows some of the intermediate results for the blueprint designs with $\xpmt{lst=1}$ and $\xpmt{lst=0}$. The objective and volume fraction convergence plots are shown in Fig.~\ref{fig:Convergenceplot}.  Steps in the plots  appear due to change in the projection parameter $\beta$. The plots have a converging nature. Next, we present a study with different $\xpmt{delrbst}$ with and without load sensitivity terms to understand their effects on the optimized designs.

\begin{figure*}
	\centering
	\begin{subfigure}[t]{0.220\textwidth}
		\centering
		$\PKcom{\xpmt{delrbst} = 0.10}$
		\includegraphics[scale=0.35]{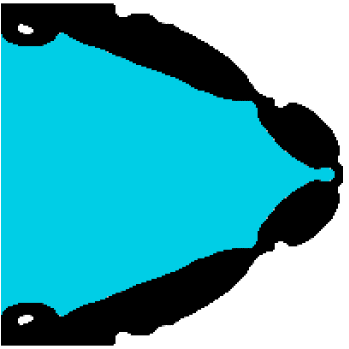}
		\caption{$u^\text{out} = -167.417$}
		\vspace*{-0.75em}
		$\xpmt{lst}=1$
		\label{fig:IN_undef_lst_1_R2}
	\end{subfigure}
	\hfill
	\begin{subfigure}[t]{0.22\textwidth}
		\centering
		$\PKcom{\xpmt{delrbst} = 0.10}$
		\includegraphics[scale=0.35]{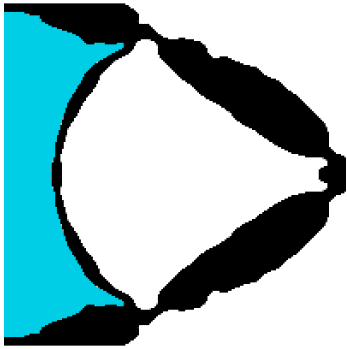}
		\caption{$u^\text{out} = -166.541$}
		\vspace*{-0.75em}
		$\xpmt{lst}=0$
		\label{fig:IN_undef_lst_0_R2}
	\end{subfigure}
	\hfill
	\begin{subfigure}[t]{0.22\textwidth}
		\centering
		$\PKcom{\xpmt{delrbst} = 0.01}$
		\includegraphics[scale=0.35]{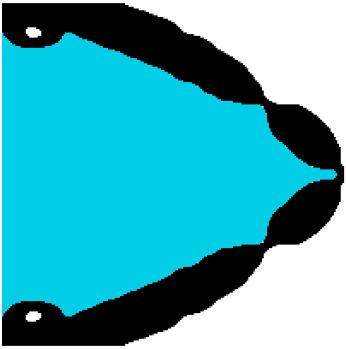}
		\caption{$u^\text{out} =  -165.024$}
		\vspace*{-0.75em}
		$\xpmt{lst}=1$
		\label{fig:IN_undef_lst_1_R5}
	\end{subfigure}
	\hfill
	\begin{subfigure}[t]{0.22\textwidth}
		\centering
		$\PKcom{\xpmt{delrbst} = 0.01}$
		\includegraphics[scale=0.35]{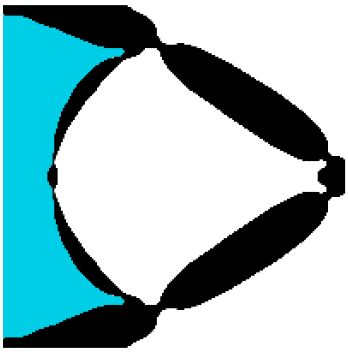}
		\caption{$u^\text{out} = -168.00$}
		\vspace*{-0.75em}
		$\xpmt{lst}=0$
		\label{fig:IN_undef_lst_0_R5}
	\end{subfigure}
	\caption{Optimized soft pneumatic inverters for different $\Delta \eta$ with $\xpmt{lst} =1$ and $\xpmt{lst=0}$.} \label{fig:IN_undefor_delrbst}
\end{figure*}

We consider $\xpmt{delrbst} = 0.10,\text{and}\,0.01$ with $\xpmt{lst=0}$, $\xpmt{lst=1}$ and the same input variables used above. Note $\xpmt{delrbst}$ with the filter radius decides the minimum length scale of the designs~\citep{kumar2022topological,fernandez2020imposing}. Lower $\xpmt{delrbst}$ indicates that the minimum length scale of the blueprint and eroded designs are close. We call $\xpmt{SoRoTop}$ with the above $\xpmt{delrbst}$ values, $\xpmt{lst}=1$ or $\xpmt{lst}=0$ and with the above parameters. The optimized results are shown in Fig.~\ref{fig:IN_undefor_delrbst}. One can note that $\xpmt{lst}$ influences the optimized designs as noted above and also in~\cite{kumar2023TOPress}. With $\xpmt{lst} =0$, the optimized designs (Fig.~\ref{fig:IN_undef_lst_0_R2} and Fig.~\ref{fig:IN_undef_lst_0_R5}) contain intermediate arch connections--that restrict the full development of the pressure chambers. However, with  $\xpmt{lst} =1$, unnecessary intermediate arch appendages are not obtained with the same parameters.   Based on the numerical results, neglecting load sensitivities in penultimate load cases may not be the right idea. We use $\xpmt{lst} =1$ for soft pneumatic actuators designed below. 

\begin{figure}
	\centering
	\includegraphics[scale=1.5]{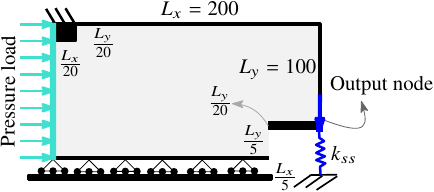}
	\caption{Gripper design domain} \label{fig:GripperDesisgn}
\end{figure}
\begin{figure}
	\centering	
	\begin{subfigure}{0.45\textwidth}
		\centering
		\includegraphics[scale=0.5]{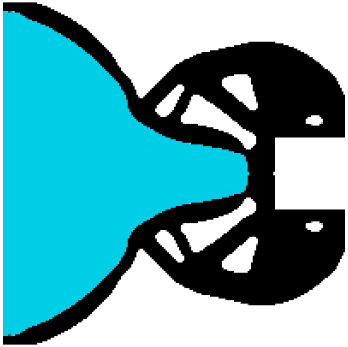}
		\caption{$ u^\text{out} = -154.479$}
		\vspace*{-0.5em}
		$\xpmt{lst}=1$ \label{fig:GP_udef_lst_1}
	\end{subfigure}
	\hfill
	\begin{subfigure}{0.45\textwidth}
		\centering
		\includegraphics[scale=0.40]{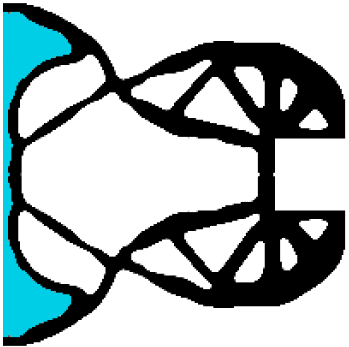}
		\caption{$ u^\text{out} =-131.151$}
		\vspace*{-0.5em}
		$\xpmt{lst}=0$ \label{fig:GP_udef_lst_0}
	\end{subfigure}
	\caption{Pneumatically actuated soft gripper mechanisms}
\end{figure}

\begin{figure}[h!]
	\centering
	\includegraphics[scale=1.5]{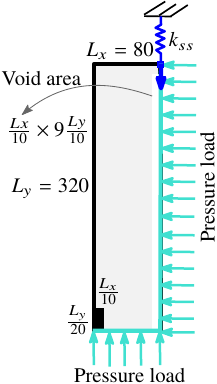}
	\caption{Soft pneumatic bending actuator design domain} \label{fig:Actuator1}
\end{figure}
\subsection{Soft pneumatic gripper}
Soft pneumatic grippers are typically designed for applications that require gripping actions, e.g., gripping fruits, vegetables, eggs, pick and place items, etc. Considering the design-dependent characteristics of the load, 2D soft pneumatic grippers are designed in~\cite{kumar2020topology,kumar2022topological,kumar2022improved,lu2021topology}, whereas Refs.~\cite{panganiban2010topology,kumar2021topology,pinskier2022automated} design 3D soft grippers.

We modify $\xpmt{SoRoTop}$ code to design a soft pneumatic gripper herein. The half-symmetric design domain is displayed in Fig.~\ref{fig:GripperDesisgn}. The pressure and displacement boundary conditions are applied as shown in Fig.~\ref{fig:GripperDesisgn}. The pressure load is applied on the left edge and and gripping action is desired at the right edge. Gripping jaw with dimension $\frac{L_x}{5}\times \frac{L_y}{20}$  exists. To facilitate an object to be placed, a void region of dimension $\frac{L_x}{5}\times \frac{L_y}{5}$ exists at the right/bottom edge. A solid non-design domain with dimension $\frac{L_x}{20} \times \frac{L_y}{20}$ also exists near the top left edge. One can modify the dimensions of the jaw and the void regions as per their requirement. 
The following modifications in $\xpmt{SoRoTop}$ are done for designing soft pneumatic gripper:
Lines~34-38 are replaced by
\begin{lstlisting}[basicstyle=\scriptsize\ttfamily,breaklines=true,numbers=none,frame=tb]
	elNrs = reshape(1:nel,nely,nelx);  % element grid
	s1=elNrs(1:nely/20,1:nelx/20);  % Elem with rho =1
	s2=elNrs(3*nely/4:4*nely/5,4*nelx/5:nelx); % Elem with rho =1
	v1=elNrs(4*nely/5:nely,4*nelx/5:nelx); % Elem with rho =0
	s = union (s1,s2); % Total elem with rho =1
	[NDS, NDV ] = deal( s, v1 );
	act = setdiff((1 : nel)', union( NDS, NDV )); %active set
	opelem = elNrs(3*nely/4,nelx); % finding the output element
	opnode = Pdofs(opelem,2); % finding the output node
	opdof = 2*opnode; % output degree of freedom
\end{lstlisting}
other parts remain as it. With above modification, $\xpmt{SoRoTop}$ is called as
\begin{lstlisting}[basicstyle=\scriptsize\ttfamily,breaklines=true,numbers=none,frame=tb]
	SoRoTop(200,100,0.30,0.90,3,5.6,1,0.1,10,lst,128,0.10,400)
\end{lstlisting}
with $\xpmt{nelx} = 200,\,\xpmt{nely} =100,\,\xpmt{volfrac} = 0.30,\,\xpmt{sefrac} = 0.90,\,\xpmt{penal} =3,\,\xpmt{rmin} = 5.6,\,\xpmt{kss}=1;\xpmt{etaf} =0.1,\,\xpmt{betaf}=10,\,\xpmt{lst}=1/0,\,\xpmt{betamax}=128,\,\xpmt{delrbst}=0.10$ and $\xpmt{maxit}=400$. The optimized results with $\xpmt{lst}=1$ and $\xpmt{lst}=0$ are shown in Fig.~\ref{fig:GP_udef_lst_1} and Fig.~\ref{fig:GP_udef_lst_0}, respectively. One again notes that the topology of the optimized designs with $\xpmt{lst}=1$ and $\xpmt{lst}=0$ are different. In the former case, a large part of the mechanism is made of a pressure chamber, whereas in the latter case, the size of the pressure chamber is relatively smaller. The output displacements of these gripper mechanisms are in the desired direction, in that the soft mechanism obtained with $\xpmt{lst}=1$ performs relatively better than that with $\xpmt{lst}=0$.
\subsection{Soft pneumatic bending actuator}
Herein, $\xpmt{SoRoTop}$ is modified to design a member of the pneumatic networks (PneuNets) mentioned in~\cite{lu2022optimal}. The PneuNet is designed to achieve a bending motion. The symmetric-half design domain of a member is shown in Fig.~\ref{fig:Actuator1}. The pressure and displacement boundary conditions are also shown in the figure. It is desired that the output node upon pneumatic actuation should move down as depicted in Fig.~\ref{fig:Actuator1}. The non-design solid domain with dimension $\frac{Lx}{10} \times \frac{Ly}{20}$ and non-design void domain of size $\frac{Lx}{10} \times 9\frac{Ly}{10}$ exist. 

The following modification in $\xpmt{SoRoTop}$ code is performed for designing this soft actuator:

Lines~34-38 are replaced by
\begin{lstlisting}[basicstyle=\scriptsize\ttfamily,breaklines=true,numbers=none,frame=tb]
	elNrs = reshape(1:nel,nely,nelx);       % element grid
	s1 = elNrs(19*nely/20:nely,1:nelx/10); % Elem with rho =1
	v1 = elNrs(nely/10:nely,9*nelx/10:nelx); % Elem with rho =0
	[NDS, NDV ] = deal( s1, v1 );
	act = setdiff((1 : nel)', union( NDS, NDV )); %active set
	opdof = 2*Tnode(end); % output degree of freedom
\end{lstlisting}
Line~41 is modified for applying the pressure load boundary condition as
\begin{lstlisting}[basicstyle=\scriptsize\ttfamily,breaklines=true,numbers=none,frame=tb]
	PF([Lnode, Tnode]) = 0; PF([Rnode, Bnode]) = Pin; % applying pressure load
\end{lstlisting}		
and line~45 is changed for applying the displacement boundary condition as
\begin{lstlisting}[basicstyle=\scriptsize\ttfamily,breaklines=true,numbers=none,frame=tb]
	fixedUdofs = [2*Lnode(end:-1:19*nely/20+1)-1  2*Lnode(end:-1:19*nely/20+1) 2*Rnode-1]; %fixed displ.
\end{lstlisting}
With the above modifications, $\xpmt{SoRoTop}$ code is called as
\begin{lstlisting}[basicstyle=\scriptsize\ttfamily,breaklines=true,numbers=none,frame=tb]
	SoRoTop(80,320,0.2,0.90,3,7.6,1,0.1,10,1,128,0.1,400)
\end{lstlisting}
with $\xpmt{nelx} = 80,\,\xpmt{nely} =320,\,\xpmt{volfrac} = 0.20,\,\xpmt{sefrac} = 0.90,\,\xpmt{penal} =3,\,\xpmt{rmin} = 7.6,\,\xpmt{kss}=1;\xpmt{etaf} =0.1,\,\xpmt{betaf}=10,\,\xpmt{lst}=1,\,\xpmt{betamax}=128,\,\xpmt{delrbst}=0.10$ and $\xpmt{maxit}=400$. As the design domain is symmetric about $y-$axis. In the plotting subroutines (Sec.~\ref{sec:post-optimization}), line~126, line~133 and line~136 are changed to
\begin{lstlisting}[basicstyle=\scriptsize\ttfamily,breaklines=true,numbers=none,frame=tb]
	X1 = 2*max(node(:,2))-reshape(node(elem(:,2:5)',2),4,nel); % for x-symmetry	
	patch(X1(:,i), Y(:,i), [1-elemP(i)],'FaceColor',[0 0.8078 0.90],'EdgeColor','none')
	patch(X1(:,i), Y(:,i), [1-elemP(i)],'FaceColor','w','EdgeColor','none')
\end{lstlisting}
respectively. Likewise, line~144, line~150 and line~153 are modified to
\begin{lstlisting}[basicstyle=\scriptsize\ttfamily,breaklines=true,numbers=none,frame=tb]
	Xn1 = 2*max(node(:,2))-reshape(xn(elem(:,[2:5])',2),4,nel); % for symmetry about x-axis
	patch(Xn1(:,i), Yn(:,i), [1-elemP(i)],'FaceColor',[0 0.8078 0.90],'EdgeColor','none')
	patch(Xn1(:,i), Yn(:,i), [1-elemP(i)],'FaceColor','w','EdgeColor','none')	
\end{lstlisting}
respectively. 
\begin{figure}
	\centering	
	\begin{subfigure}[t]{0.45\textwidth}
		\centering
		\includegraphics[scale=0.5]{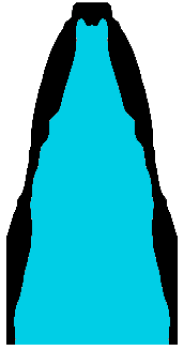}
		\caption{$ u^\text{out} = -586.542$} \label{fig:Act_1_lst_1}
	\end{subfigure}
	\hfill
	\begin{subfigure}[t]{0.45\textwidth}
		\centering
		\includegraphics[scale=0.5]{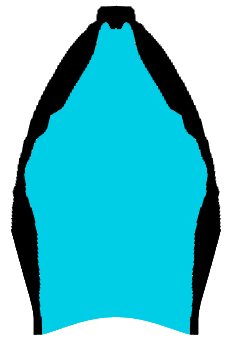}
		\caption{Deformed profile} \label{fig:Act_1_def_lst_1}
	\end{subfigure}
	\caption{Optimized soft pneumatic bending actuator. (\subref{fig:Act_1_lst_1}) Optimized design and (\subref{fig:Act_1_def_lst_1}) Scaled deformed design}
\end{figure}

Figure~\ref{fig:Act_1_lst_1} display the full optimized result with the final pressure field. The optimized design (Fig.~\ref{fig:Act_1_lst_1}) and its deformation profile (Fig.~\ref{fig:Act_1_def_lst_1}) resemble that obtained in \cite{lu2022optimal}.  Therefore, the performance  of the PneuNets made via the optimized result shown in Fig.~\ref{fig:Act_1_lst_1} is expected as that of \cite{lu2022optimal}.

\subsection{Centrally pressurized soft pneumatic actuator}
\begin{figure}
	\centering
	\includegraphics[scale=1.5]{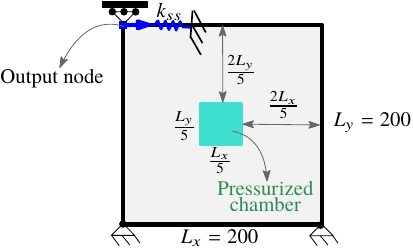}
	\caption{Centrally pressurized soft pneumatic actuator design domain} \label{fig:centralactuator}
\end{figure}

Next, to demonstrate the additional capabilities of the presented code, we design a centrally pressurized soft pneumatic actuator. The design domain of the actuator is depicted in Fig.~\ref{fig:centralactuator}. The pressure load is applied in the central chamber having dimension $\frac{L_x}{10} \times \frac{L_y}{10}$, which is considered a void region. Edges of the domain are set at zero pressure level. The output location and the desired deformation are depicted in the figure. The left and right ends of the bottom edge are fixed (Fig.~\ref{fig:centralactuator}). 
The following modifications are done to optimize the centrally pressurized soft pneumatic actuator. Lines~35-36 are replaced by
\begin{lstlisting}[basicstyle=\scriptsize\ttfamily,breaklines=true,numbers=none,frame=tb]
	v1 = elNrs(2*nely/5:3*nely/5,2*nelx/5:3*nelx/5);
	[NDS, NDV ] = deal([], v1);
\end{lstlisting}
As the desired output deformation is in positive $x-$direction, the line~38 is replaced by
\begin{lstlisting}[basicstyle=\scriptsize\ttfamily,breaklines=true,numbers=none,frame=tb]
	opdof = 2*Tnode(1)-1; % output degree of freedom
\end{lstlisting}
The pressure is applied on the central region (Fig.~\ref{fig:centralactuator}); thus, the nodes associated with the domain are determined and stored in vector $\xpmt{nodeV1}$. A line is inserted after line~38 for  $\xpmt{nodeV1}$ as
\begin{lstlisting}[basicstyle=\scriptsize\ttfamily,breaklines=true,numbers=none,frame=tb]
	nodeV1 = unique(Pdofs(v,:)); % pressurized nodes
\end{lstlisting}
The given pressure load conditions are applied by altering line~41 to
\begin{lstlisting}[basicstyle=\scriptsize\ttfamily,breaklines=true,numbers=none,frame=tb]
	PF([Lnode, Bnode,Tnode, Rnode]) = 0; PF(nodeV1) = Pin;
\end{lstlisting}
Line~45 is changed to apply the displacement boundary conditions as
\begin{lstlisting}[basicstyle=\scriptsize\ttfamily,breaklines=true,numbers=none,frame=tb]
	fixedUdofs = [2*Bnode(1)-1  2*Bnode(1)  2*Bnode(end)-1 2*Bnode(end) 2*Tnode(1)]; %fixed displ.	
\end{lstlisting}
In addition to the above mentioned changes, we remove the symmetry plotting lines from PART~7 and PART~8. Finally, we call the code as
\begin{lstlisting}[basicstyle=\scriptsize\ttfamily,breaklines=true,numbers=none,frame=tb]
	SoRoTop(200,200,0.25,0.9,3,6.6,1,0.1,10,1,128,0.05,400)
\end{lstlisting}
with $\xpmt{nelx} = 200,\,\xpmt{nely} =200,\,\xpmt{volfrac} = 0.25,\,\xpmt{sefrac} = 0.90,\,\xpmt{penal} =3,\,\xpmt{rmin} = 6.6,\,\xpmt{kss}=1;\xpmt{etaf} =0.1,\,\xpmt{betaf}=10,\,\xpmt{lst}=1,\,\xpmt{betamax}=128,\,\xpmt{delrbst}=00.1$ and $\xpmt{maxit}=400$. 

\begin{figure}
	\centering	
	\begin{subfigure}[t]{0.45\textwidth}
		\centering
		\includegraphics[scale=0.5]{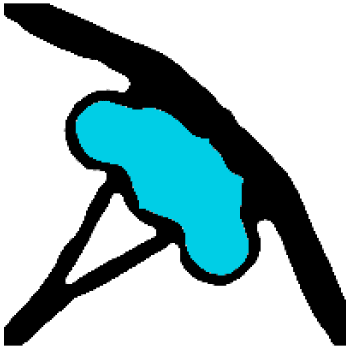}
		\caption{$ u^\text{out} = 100.7640$} \label{fig:central_1_lst_1}
	\end{subfigure}
	\hfill
	\begin{subfigure}[t]{0.45\textwidth}
		\centering
		\includegraphics[scale=0.5]{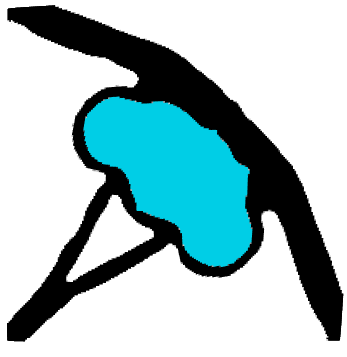}
		\caption{Deformed profile} \label{fig:central_1_def_lst_1}
	\end{subfigure}
	\caption{Optimized centrally pressurized soft pneumatic actuator. (\subref{fig:Act_1_lst_1}) Optimized design and (\subref{fig:Act_1_def_lst_1}) Scaled deformed design}
\end{figure}	
Figure~\ref{fig:central_1_lst_1} provides the optimized design for the centrally pressurized soft pneumatic actuator. The pressure chamber gets an arbitrary shape that helps achieve the desired deformation, as noted in the deformed profile shown in Fig.~\ref{fig:central_1_def_lst_1}. 

Likewise, a user can optimize different pneumatically actuated soft robots by changing the displacement and pressure boundary conditions using  $\xpmt{SoRoTop}$. In addition, the code can readily be extended with different necessary constraints if needed.
\section{Concluding remarks}\label{sec:conclusions}
This paper presents a MATLAB code, named $\xpmt{SoRoTop}$,  for designing soft pneumatic actuators using TO. Such actuators find various applications; however, it is a challenging and complex task to model the design-dependent nature of the pneumatic loads within a TO setting, as such loads change their direction, magnitude, and/or location with the optimization iterations. Thus, to ease newcomers' and students' learning paths toward designing soft pneumatic actuators, the code is developed per the method first introduced in~\cite{kumar2020topology}. The code uses the robust formulation with eroded and blueprint designs to circumvent the appearance of single-node connections in the optimized mechanisms.  $\xpmt{SoRoTop}$ is explained in detail and is extended to optimize different soft pneumatic actuators.  The paper also demonstrates that load sensitivities have a significance impact on the optimized topology of these actuators. Hence, it is essential not to overlook them when conducting a TO.

The method proposed in~\cite{kumar2020topology} is employed to model the design-dependent pneumatic load. The flow and drainage terms are interpolated using the Heaviside functions. The obtained pressure field is converted to consistent nodal loads using a transformation matrix. The objective's and constraints' sensitivities are determined using the adjoint-variable method; thus, the load sensitivities. One can toggle load sensitivity terms to 1 or 0 using $\xpmt{lst}$ input parameter. The optimization is formulated as a min-max problem, involving both blueprint and eroded design descriptions. The method of moving asymptotes is employed for the optimization. Volume constraint is applied to the blueprint design, whereas a strain energy constraint is applied to the eroded design. A strain energy fraction parameter $\xpmt{sefrac}$ is introduced to apply the latter constraint which assists in attaining optimized designs that can endure the applied load without sacrificing performance. Based on the results of the numerical experiments performed, we observe, $\xpmt{sefrac}\in [0.80\,\, 0.95]$ performs well, resulting in achievable, leak-proof optimized designs. One may also opt to reduce $\xpmt{sefrac}$ in a specific problem to ensure a leak-proof design. Objective and volume fraction plots reveal that they have a converging nature. The volume fraction remains active and satisfied at the end of the optimization.   

$\xpmt{SoRoTop}$ is provided in Appendix~\ref{sec:sorotop}, and it is extended to optimize different soft pneumatic actuators. The code is structured into pre-optimization, MMA optimization, and post-optimization operations, and each operation is described in detail. The design of four pneumatically activated soft robots demonstrates the efficacy and robustness of code. Integrating the code with nonlinear mechanics forms a complex and exciting future direction. We believe that newcomers, students, and the soft robotics research community will take advantage of the provided code, and they will utilize and extend the code to generate soft pneumatic actuators for different applications.

\section*{Acknowledgments}
The author thanks Matthijs Langelaar and  Ole Sigmund for discussions on the method in the past, Krister Svanberg (krille@math.kth.se) for providing MATLAB codes of the MMA optimizer, and Indian Institute of Technology Hyderabad for the support under the seed grant with project file number SG/IITH/F297/2022-23/SG-155.

\section*{Declaration of Competing Interest}
None.
\begin{appendices}
	\numberwithin{equation}{section}
	\numberwithin{figure}{section}
	\onecolumn
	\section{The MATLAB code: \texttt{SoRoTop}}\label{sec:sorotop}
	\begin{lstlisting}[basicstyle=\scriptsize\ttfamily,breaklines=true]
function SoRoTop(nelx,nely,volfrac,sefrac,penal,rmin,kss,etaf,betaf,lst,betamax,delrbst,maxit)
%% ___PART 1.____________________________MATERIAL AND FLOW PARAMETERS
E1 = 1;
Emin = E1*1e-6;
nu = 0.30;
[Kv,epsf,r,Dels] = deal(1,1e-7,0.1,2);                       % flow parameters
[Ds, Kvs]= deal((log(r)/Dels)^2*epsf,Kv*(1 - epsf));  % flow parameters
%% ____PART 2._______________FINITE ELEMENT and NON-DESIGN DOMAIN PREPARATION
[nel,nno] = deal(nelx*nely, (nelx+1)*(nely+1));
nodenrs = reshape(1:(1+nelx)*(1+nely),1+nely,1+nelx);
edofVec = reshape(2*nodenrs(1:end-1,1:end-1)+1,nelx*nely,1);
Udofs = repmat(edofVec,1,8)+repmat([0 1 2*nely+[2 3 0 1] -2 -1],nelx*nely,1);
[Lnode,Rnode]= deal(1:nely+1, (nno-nely):nno);
[Bnode,Tnode]= deal((nely+1):(nely+1):nno, 1:(nely+1):(nno-nely));
[Pdofs,allPdofs,allUdofs] = deal(Udofs(:,2:2:end)/2,1:nno,1:2*nno);
Kp = 1/6*[4 -1 -2 -1;-1 4 -1 -2; -2 -1 4 -1; -1 -2 -1 4]; % flow matrix: Darcy Law
KDp = 1/36*[4 2 1 2; 2 4 2 1; 1 2 4 2; 2 1 2 4]; % Drainage matrix
Te = 1/12*[-2 2 1 -1;-2 -1 1 2;-2 2 1 -1;-1 -2 2 1;-1 1 2 -2; -1 -2 2 1; -1 1 2 -2; -2 -1 1 2]; % transformation matrix
A11 = [12  3 -6 -3;  3 12  3  0; -6  3 12 -3; -3  0 -3 12];
A12 = [-6 -3  0  3; -3 -6 -3 -6;  0 -3 -6  3;  3 -6  3 -6];
B11 = [-4  3 -2  9;  3 -4 -9  4; -2 -9 -4 -3;  9  4 -3 -4];
B12 = [ 2 -3  4 -9; -3  2  9 -2;  4  9  2  3; -9 -2  3  2];
Ke = 1/(1-nu^2)/24*([A11 A12;A12' A11]+nu*[B11 B12;B12' B11]);             %stiffness matrix
iP = reshape(kron(Pdofs,ones(4,1))',16*nel,1);
jP = reshape(kron(Pdofs,ones(1,4))',16*nel,1);
iT = reshape(kron(Udofs,ones(4,1))',32*nel,1);
jT = reshape(kron(Pdofs,ones(1,8))',32*nel,1);
iK = reshape(kron(Udofs,ones(8,1))',64*nel,1);
jK = reshape(kron(Udofs,ones(1,8))',64*nel,1);
IFprj=@(xv,etaf,betaf)((tanh(betaf*etaf) + tanh(betaf*(xv-etaf)))/...            %projection function
(tanh(betaf*etaf) + tanh(betaf*(1 - etaf))));
dIFprj=@(xv,etaf,betaf) betaf*(1-tanh(betaf*(xv-etaf)).^2)...
/(tanh(betaf*etaf)+tanh(betaf*(1-etaf)));                    % derivative of the projection function
elNrs = reshape(1:nel,nely,nelx);                                    % element grid
s1 = elNrs(1:nely/20,1:nelx/20);                                      % Solid element or element with rho =1
[NDS, NDV ] = deal( s1, [] );
act = setdiff((1 : nel)', union( NDS, NDV ));
opdof = 2*Bnode(end)-1;
%% ____PART 3.__PRESSURE & STRUCTURE B.C's, LOADs, DOFs, Lag. Multi.,sL
[PF, Pin] =deal(0.00001*ones(nno,1),1); %pressure-field preparation
PF([Tnode, Rnode]) = 0; PF(Lnode) = Pin; % applying pressure load
fixedPdofs = allPdofs(PF~=0.00001);
freePdofs  = setdiff(allPdofs,fixedPdofs);
pfixeddofsv = [fixedPdofs' PF(fixedPdofs)]; % p-fixed and its value
fixedUdofs = [2*Tnode(1:nely/20)-1  2*Tnode(1:nely/20)  2*Bnode]; %fixed displ.
freeUdofs = setdiff(allUdofs,fixedUdofs);
[L,U,lam2] = deal(zeros(2*nno,1));
[lam1,mu1] = deal(zeros(nno,1)); %initialize lambda
[L(opdof)] = 1 ;                  % dummy load and spring constant
%% ___PART 4._________________________________________FILTER PREPARATION
iH = ones(nelx*nely*(2*(ceil(rmin)-1)+1)^2,1);
jH = ones(size(iH));
sH = zeros(size(iH));
k = 0;
for i1 = 1:nelx
for j1 = 1:nely
e1 = (i1-1)*nely+j1;
for i2 = max(i1-(ceil(rmin)-1),1):min(i1+(ceil(rmin)-1),nelx)
for j2 = max(j1-(ceil(rmin)-1),1):min(j1+(ceil(rmin)-1),nely)
e2 = (i2-1)*nely+j2;
k = k+1;
iH(k) = e1;
jH(k) = e2;
sH(k) = max(0,rmin-sqrt((i1-i2)^2+(j1-j2)^2));
end
end
end
end
H = sparse(iH,jH,sH);
Hs = H./sum(H,2);                             % matrix of weights (filter)
%% ___PART 5.__________________________MMA OPTIMIZATION PREPARATION & INITIALIZATION
x = zeros(nel,1); % design variable
x(act) = (volfrac*(nel-length(NDV))-length(NDS) )/length(act); x(NDS) = 1;
[nMMA,pMMA,qMMA] = deal(length(act),2,2);
[mMMA,xMMA,xTilde,mvLt] = deal(pMMA+qMMA,x(act),x,0.1);
[xminvec,xmaxvec] = deal(zeros(nMMA,1),ones(nMMA,1)); %Min. & Max
[low, upp] = deal(xminvec,xmaxvec); % Low and Upp limits MMA
[cMMA,dMMA, a0] = deal(1000*ones(mMMA,1),zeros(mMMA,1),1);
aMMA = [ones(pMMA,1); zeros(qMMA,1)];
[xold1,xold2] = deal(xMMA);
[betap,loop, change] = deal(1,0,1);
[etab,etae] =deal(0.5,0.5+delrbst);
costadd = 10000;
%% ____PART 6_____________________________________MMA OPTIMIZATION LOOP
while(loop<maxit && change>0.0001)
loop = loop + 1;  % Updating the opt. iteration
%___PART 6.1___________Compute blueprint and eroded physical desing variables
xTilde(act) = xMMA;  xTilde =Hs'*xTilde; xTilde(NDS)=1;  xTilde(NDV)=0;
xphysb  =  IFprj(xTilde,etab,betap); xphysb(NDS)=1; xphysb(NDV)=0;
xphyse = IFprj(xTilde,etae,betap); xphyse(NDS)=1; xphyse(NDV)=0;
%___PART 6.2__________Performing blueprint design analysis using ObjObjSens function
[objb, objsensb,volb,volsensb, ~, ~,Ub,Fb,PFb] = ObjCnst_ObjCnst_Sens(xphysb,nel,E1,Emin,penal,Kv,Kvs,epsf,Ds,etaf,betaf,Udofs,freeUdofs,
Pdofs,pfixeddofsv,fixedPdofs,freePdofs,iP,jP,iT,jT,iK,jK,Kp,KDp,Te,Ke,opdof,kss,loop,IFprj,dIFprj,L,U,lam1,lam2,mu1,lst,volfrac,sefrac);
%___PART 6.3___________Performing eroded design analysis using ObjObjSens function
[obje, objsense,~,~, SEe, SEsense,~,~,~] = ObjCnst_ObjCnst_Sens(xphyse,nel,E1, Emin,penal,Kv,Kvs,epsf,Ds,etaf,betaf,Udofs,freeUdofs, ...
Pdofs,pfixeddofsv,fixedPdofs,freePdofs,iP,jP,iT,jT,iK,jK,Kp,KDp,Te,Ke,opdof,kss,loop,IFprj,dIFprj,L,U,lam1,lam2,mu1,lst,volfrac,sefrac);
%___PART 6.4_____________________Filtering and projecting objective and constraints sensitivities
objsensb = Hs'*(objsensb.*dIFprj(xphysb,etab,betap)); % blueprint sensitiivty
objsense= Hs'*(objsense.*dIFprj(xphyse,etae,betap));
volsensb = Hs'*(volsensb.*dIFprj(xphysb,etab,betap));
SEsense = Hs'*(SEsense.*dIFprj(xphyse,etae,betap));
%___PART 6.5_________________Stacking constraints and their sensitivities
constr =[volb SEe];
constrsens = [volsensb SEsense];
normf = 1;
%___PART 6.6______________________SETTING and CALLING MMA OPTIMIZATION
fval = [costadd + objb*normf,costadd + obje*normf,constr]';
dfdx  = [objsensb(act,:)*normf, objsense(act,:)*normf, constrsens(act,:)]';
[xminvec, xmaxvec]= deal(max(0, xMMA - mvLt),min(1, xMMA + mvLt));
[xmma,~,~,~,~,~,~,~,~,low,upp] = mmasub(mMMA,nMMA,loop,xMMA,xminvec,xmaxvec,xold1,xold2, ...
0,0,0,fval,dfdx,0*dfdx,low,upp,a0,aMMA,cMMA,dMMA);
%___PART 6.7____________Updating__________
[xold2,xold1, xnew]= deal(xold1, xMMA,xmma);
change = max(abs(xnew-xMMA)); % Calculating change
xMMA = xnew;
if(mod(loop,50)==0 && betap<=betamax), betap= betap*2;end % beta updation
%___PART 6.8____________________________Printing and plotting results
fprintf(' It.:%5i Obji.:%11.4f  Obje.:%11.4f Voli.:%7.3f ch.:%7.3f\n',loop,fval(1),fval(2),mean(xphysb),change);
colormap(gray); imagesc(1-reshape(xphysb, nely, nelx));caxis([0 1]);axis equal off;drawnow;
end
%% ______PART 7____plotting results with final pressure field___________
PFP = figure(2); set(PFP,'color','w'); axis equal off, hold on; colormap('gray');
node = [ (1:nno)' reshape(repmat(0:nelx,nely+1,1),nno,1) repmat(0:-1:-nely,1, nelx+1)']; % nodal coordinates
elem(:,1) = (1:nel)'; elem(:,2:5) = Pdofs; % element and connectivity information
X = reshape(node(elem(:,2:5)',2),4,nel); Y = reshape(node(elem(:,2:5)',3),4,nel);
Y1 = 2*min(node(:,3))-reshape(node(elem(:,2:5)',3),4,nel); % for x-symmetry
for i = 1:nel,elemP(i) = sum(PFb(elem(i,2:5)))/4/Pin;end
patch(X, Y,  [1-xphysb],'EdgeColor','none');caxis([0 1]);
patch(X, Y1, [1-xphysb],'EdgeColor','none');caxis([0 1]);
for i = 1:nel
if (xphysb(i)<0.2 && elemP(i)>0.70)
patch(X(:,i), Y(:,i), [1-elemP(i)],'FaceColor',[0 0.8078 0.90],'EdgeColor','none')
patch(X(:,i), Y1(:,i), [1-elemP(i)],'FaceColor',[0 0.8078 0.90],'EdgeColor','none')
elseif (xphysb(i)<0.2 && elemP(i)<0.70)
patch(X(:,i), Y(:,i), [1-elemP(i)],'FaceColor','w','EdgeColor','none')
patch(X(:,i), Y1(:,i), [1-elemP(i)],'FaceColor','w','EdgeColor','none')
end
end
%% ______PART 8____Plotting deformed profile____________________________
DFP = figure(3); set(DFP,'color','w'); axis equal off, hold on;colormap(gray);caxis([0 1]);
xn= node;                             % defomed nodal position
xn(:,2) = node(:,2) + 0.0025*Ub(1:2:end); xn(:,3) = node(:,3) + 0.0025*Ub(2:2:end);
Xn = reshape(xn(elem(:,[2:5])',2),4,nel); Yn = reshape(xn(elem(:,[2:5])',3),4,nel);
Yn1 = 2*min(node(:,3))-reshape(xn(elem(:,[2:5])',3),4,nel); % for symmetry about x-axis
patch(Xn, Yn,   [1-xphysb],'EdgeColor','none');
patch(Xn, Yn1,  [1-xphysb],'EdgeColor','none');
for i = 1:nel
if (xphysb(i)<0.2 && elemP(i)>0.70)
patch(Xn(:,i), Yn(:,i), [1-elemP(i)],'FaceColor',[0 0.8078 0.90],'EdgeColor','none');
patch(Xn(:,i), Yn1(:,i), [1-elemP(i)],'FaceColor',[0 0.8078 0.90],'EdgeColor','none');
elseif (xphysb(i)<0.2 && elemP(i)<0.70)
patch(Xn(:,i), Yn(:,i), [1-elemP(i)],'FaceColor','w','EdgeColor','none');
patch(Xn(:,i), Yn1(:,i), [1-elemP(i)],'FaceColor','w','EdgeColor','none');
end
end
%% ___________PART 9_______ Analyses function
function[obj,objsens,vol,volsens, SEc, SEsens,U,F,PF] =ObjCnst_ObjCnst_Sens(xphys,nel,E1,Emin,penal,Kv,kvs,epsf,Ds,etaf,betaf,Udofs,...
freeUdofs,Pdofs,pfixeddofsv,fixedPdofs,freePdofs,iP,jP,iT,jT,iK,jK,Kp,KDp,Te,Ke,opdof,kss,loop,IFprj,dIFprj,L,U,lam1,lam2,mu1,lst,volfrac,sefrac)
% ___PATT 9.1_______SOLVING FLOW BALANCE EQUATION
Kc = Kv*(1-(1-epsf)*IFprj(xphys,etaf,betaf));         %Flow coefficient
Dc = Ds*IFprj(xphys,etaf,betaf);                            %Drainage coefficient
Ae = reshape(Kp(:)*Kc' + KDp(:)*Dc',16*nel,1);    %Elemental flow matrix in vector form
AG = (sparse(iP,jP,Ae)+ sparse(iP,jP,Ae)')/2;         %Global flow matrix
Aff = AG(freePdofs,freePdofs);     %AG for free pressure dofs
dAff_ldl = decomposition(Aff,'ldl'); % Decomposing Aff matrix
PF(freePdofs,1) = dAff_ldl\(-AG(freePdofs,fixedPdofs)*pfixeddofsv(:,2));
PF(pfixeddofsv(:,1),1) = pfixeddofsv(:,2);              % Final P-field
%__PART 9.2_DETERMINING CONSISTENT NODAL LOADS and GLOBAL Disp. Vector
Ts = reshape(Te(:)*ones(1,nel), 32*nel, 1);        %Elemental transformation matrix in vector form
TG = sparse(iT, jT, Ts);                                       %Global transformation matrix
F = -TG*PF;                                                       % Dertmining nodal forces
E = Emin + xphys.^penal*(E1 - Emin);                %Material interpolation
Ks = reshape(Ke(:)*E',64*nel,1);                         %Elemental stiffness matrix in vector form
KG = (sparse(iK,jK,Ks) + sparse(iK,jK,Ks)')/2;    %Global stiffnes matrix
KG(opdof,opdof) = KG(opdof,opdof) + kss;          % adding the workpiece stiffness
dKG_chol = decomposition(KG(freeUdofs,freeUdofs),'chol','lower'); % decomposed freedofs stiffness
U(freeUdofs) = dKG_chol\F(freeUdofs); %Global Disp. Vect.
%__PART 9.3______objective evaluation
obj = L'*U; % maximizing the output deformation
%__PART 9.4__________sensitivity analysis
lam2(freeUdofs) = -dKG_chol\L(freeUdofs);
lam1(freePdofs) = -(lam2(freeUdofs)'*TG(freeUdofs,freePdofs))/dAff_ldl;
objsT1 = (E1 - Emin)*penal*xphys.^(penal - 1).*sum((lam2(Udofs)*Ke).*U(Udofs),2);
dC1k = -dIFprj(xphys,etaf,betaf).* sum((lam1(Pdofs)*(kvs*Kp)) .* PF(Pdofs),2);
dC1d =  dIFprj(xphys,etaf,betaf).* sum((lam1(Pdofs)*(Ds*KDp)) .* PF(Pdofs),2);
objsT2 = dC1k + dC1d;
objsens = (objsT1 + lst*objsT2); % final sensitivities
%__PART 9.5____volume sensitivities
vol = sum(xphys)/(nel*volfrac)-1;
volsens = 1/(volfrac*nel)*ones(nel,1);
%___PART 9.6____Strain energy sensitivities
if(loop==1) , SE_perm = sefrac*(0.5*U'*KG*U); save SE_perm SE_perm; end; load SE_perm
SEc =   0.5*U'*KG*U/SE_perm -1;
SET1 = -0.5*(E1 - Emin)*penal*xphys.^(penal - 1).*sum(([U(Udofs)]*Ke).*[U(Udofs)],2);
mu1(freePdofs) =   (U(freeUdofs)'*TG(freeUdofs,freePdofs))/dAff_ldl;
dSEk = -dIFprj(xphys,etaf,betaf).* sum((mu1(Pdofs)*(kvs*Kp)) .* PF(Pdofs),2);
dSEd =  dIFprj(xphys,etaf,betaf).* sum((mu1(Pdofs)*(Ds*KDp)) .* PF(Pdofs),2);
SET2 = dSEk + dSEd;
SEsens = (SET1 + SET2)/SE_perm;
\end{lstlisting}
\lstdefinestyle{nonumbers}
{numbers=none}
\begin{lstlisting}[style=nonumbers]
%%%%%%%%%%%%%%%%%%%%%%%%%%%%%%%%%%%%%%%%%%%%%%%%%%%%%%%%%%%%%%%%%%%%%%%%%%%%%%%%%%%%
%    SoRoTop is written for  pedagogical purposes. A  detailed description can be  %
%    found in the paper:"SoRoTop: a hitchhiker's guide to topology optimization    % 
%    MATLAB code for design-dependent pneumatic-driven soft robots" Optimization   % 
%    and Engineering, 2023.                                                        %
%                                                                                  %
%    Code and its extensions are available  online as supplementary material       %  
%    of the paper and also available at:                                           %
%                                      https://github.com/PrabhatIn/SoRoTop        %
%                                                                                  %
%    Please send your comment to: pkumar@mae.iith.ac.in                            %
%                                                                                  %
%    One may also refer to the following two papers for more detail:               % 
%                                                                                  %
%    1. Kumar P, Frouws JS, Langelaar M (2020) Topology optimization of fluidic    %
%    pressure-loaded structures and compliant mechanisms using the Darcy method.   %
%    Structural and Multidisciplinary Optimization 61(4):1637-1655                 %
%    2. Kumar P, Langelaar M (2021) On topology optimization of design-dependent   % 
%    pressure-loaded three-dimensional structures and compliant mechanisms.        %
%    International Journal for Numerical Methods in Engineering 122(9):2205-2220   %
%    3. P. Kumar (2023) TOPress: a MATLAB implementation for topology optimization %
%    of structures subjected to desig-dependent pressure loads, Structural and     %
%    Multidisciplinary Optimization, 66(4), 2023                                   %
%                                                                                  %   
%                                                                                  %
%                                                                                  %
%    Disclaimer:                                                                   %
%    The author does not guarantee that the code is free from erros but reserves   %
%    all rights. Further, the author shall not be liable in any event caused by    % 
%    use of the above code and its extensions                                      %
%                                                                                  %
%%%%%%%%%%%%%%%%%%%%%%%%%%%%%%%%%%%%%%%%%%%%%%%%%%%%%%%%%%%%%%%%%%%%%%%%%%%%%%%%%%%%
\end{lstlisting}

\end{appendices}



\end{document}